\documentstyle[preprint,tighten,eqsecnum,floats,epsfig,aps]{revtex}
\newcommand{\mc}{\multicolumn}
\newcommand{\lsim}{\mathrel{\mathop{\kern 0pt \rlap
  {\raise.2ex\hbox{$<$}}}
  \lower.9ex\hbox{\kern-.190em $\sim$}}}
\newcommand{\gsim}{\mathrel{\mathop{\kern 0pt \rlap
  {\raise.2ex\hbox{$>$}}}
  \lower.9ex\hbox{\kern-.190em $\sim$}}}

\begin{document}
\draft
\include{def}
\def\v#1{\mbox{\boldmath$#1$}}
\def\ket#1{|#1 \rangle}
\def\bra#1{\langle #1|}

%\title{Single and double coincidence nucleon spectra in the non--mesonic
%weak decay of $\Lambda$--hypernuclei: LDA vs Finite Nucleus
%(another one?)}

\title{Single and double coincidence nucleon spectra in the
weak decay of $\Lambda$--hypernuclei}

\author{E.~Bauer$^1$, G.~Garbarino$^2$, A. Parre\~no$^3$ and A. Ramos$^3$}

\address{$^1$Departamento de F\'{\i}sica, Universidad Nacional de
%\affiliation{$^1$Departamento de F\'{\i}sica, Universidad Nacional de
La Plata,\\ C. C. 67, 1900 La Plata, Argentina}

\address{$^2$Dipartimento di Fisica Teorica, Universit\`a di Torino and INFN, Sezione
%\affiliation{$^2$Dipartimento di Fisica Teorica, Universit\`a di Torino and INFN, Sezione
di Torino, \\I--10125 Torino, Italy}

\address{$^3$Departament d'Estructura i Constituents de la Mat\`{e}ria,
%\affiliation{$^3$Departament d'Estructura i Constituents de la Mat\`{e}ria,
Universitat de Barcelona, \\ E--08028 Barcelona, Spain}

\date{\today}
\maketitle

\begin{abstract}
Recent progress has been experienced in the field of hypernuclear
weak decay, especially concerning the ratio of the neutron-- to
proton--induced $\Lambda$ non--mesonic decay rates, $\Gamma_n/\Gamma_p$.
Theoretical analyses of nucleon
coincidence data have been performed in a finite nucleus framework.
They led to the extraction of $\Gamma_n/\Gamma_p$ values in agreement with
pure theoretical estimates, thus providing an evidence for the solution
of a longstanding puzzle.
Here we present an alternative approach to the problem, based on a nuclear matter
formalism extended to finite nuclei via the local density approximation.
The work is motivated by the exigence to
make the determination of $\Gamma_n/\Gamma_p$ from data less model dependent.
One--meson--exchange potentials are used for describing both the one-- and two--nucleon
induced decays, $\Lambda N\to nN$ and $\Lambda NN\to nNN$.
For the latter, treated within a microscopic approach, the channels $\Lambda nn\to nnn$ and
$\Lambda pp \to npp$ are included in addition to the mode $\Lambda np\to nnp$ already
considered, in a phenomenological way, in previous studies. The propagation of the final
nucleons in the residual nucleus is simulated by an intranuclear cascade
code. We evaluate single and double coincidence nucleon spectra for the non--mesonic
decay of $^{12}_\Lambda$C.
Through the comparison of our predictions with KEK coincidence data
we determine $\Gamma_n/\Gamma_p=0.43\pm 0.10$ for this hypernucleus, confirming
previous finite nucleus analyses.
The use of a high nucleon kinetic energy detection threshold such as $60$ MeV
makes the contribution of two--nucleon induced channels quite small, but final
state interaction effects are still important when extracting
$\Gamma_n/\Gamma_p$ from measured single nucleon distributions. 
Coincidence spectra suffer less from these effects even for a moderate
detection threshold such as $30$ MeV. In any case, final state
interactions have to be considered
for meaningful determinations of $\Gamma_n/\Gamma_p$.
%However, a proper evaluation of FSI
%reveals to be essential when determining $\Gamma_n/\Gamma_p$ from data.
\end{abstract}
%\pacs{PACS numbers: 21.80.+a, 13.30.Eg, 13.75.Ev}

%\maketitle

\newpage
\pagestyle{plain}
\baselineskip 16pt\vskip 48pt

\newpage
%\begin{multicols}{2}

\date{\today}

%%%%%%%%%%%%%%%%%%%%%%%%%%%%%%%%%%%%%%%%%%%%%%%%%%%%%%%
\section{Introduction}
%%%%%%%%%%%%%%%%%%%%%%%%%%%%%%%%%%%%%%%%%%%%%%%%%%%%%%%
\label{intro}
Diversified efforts have been devoted to the study
of hypernuclear weak decay in the latest years.
Theoretical reviews on the subject can be found in Refs.~\cite{Al02,Ra98} 
and recent, related experiments in
Refs.~\cite{Ha01,Sa05,Ki02,Ok04,bhang,OutaVa,Kang05,Kim06}.
Forthcoming data are expected from FINUDA \cite{fi}, while new
experiments are planned at J--PARC \cite{jparc} and HypHI \cite{hyphi}.
Strong evidences for a solution of the longstanding problem on the ratio
$\Gamma_n/\Gamma_p$ between the widths for the weak processes $\Lambda n\to nn$
and $\Lambda p \to np$ originated from theoretical analyses
\cite{prl,prc} of KEK nucleon coincidence data \cite{bhang,OutaVa,Kang05,Kim06}.
In the authors opinion, this puzzle was due to the non--trivial interpretation
of experimental data, which required
a careful analysis of nuclear medium effects on the weak decay nucleons, rather than
to a poor understanding of the weak decay mechanism.
Indeed, the lately extracted $\Gamma_n/\Gamma_p$ values \cite{prl,prc}
%(around $0.3\div 0.4$ for both $^5_\Lambda$He and $^{12}_\Lambda$C)
turned out to be in agreement with the previous, pure theoretical estimates of
Refs.~\cite{Os01,Pa02,It02,Ba03,Ba04} obtained by using one--meson--exchange potentials to
describe the one--nucleon induced, $\Lambda N\to nN$ weak transitions.

Nevertheless, further theoretical and experimental work is desirable
in order to confirm the previously mentioned evidence in favor of a solution of the
$\Gamma_n/\Gamma_p$ puzzle.
%and then have a more complete understanding of hypernuclear non--mesonic weak decay.
Indeed, on the one hand such an evidence relies on
particular theoretical descriptions of both the weak decay
mechanism and the subsequent propagation
of the produced nucleons within the residual nucleus. In this direction, the use of
alternative weak decay or/and intranuclear cascade models is of interest.
On the other hand, one has to consider that another problem of the field is still
unsolved: it concerns the asymmetry of the protons emitted in the non--mesonic
decay of polarized hypernuclei, measured to be not far from zero in recent
experiments \cite{Aj00,Ma04} while a large negative number is predicted by the theoretical
models\cite{Pa02,asyth,barbero,oka,assum}. Recently, a strong effect of nucleon final
state interactions (FSI) was pointed out \cite{asyth} without,
however, bringing new hints for a possible solution of the asymmetry puzzle.
The connections existing among the weak decay observables
[$\Gamma_n/\Gamma_p$, $\Gamma_{\it NM}=\Gamma_n+\Gamma_p+\Gamma_2$, $\Gamma_2$
being the width for two--nucleon induced decays, $\Lambda NN\to nNN$,
and the asymmetry parameters] and the question concerning the validity
of the $\Delta I=1/2$ isospin rule in the non--mesonic decay
is another important issue which deserves future investigations in the prospect
of a better understanding of baryon--baryon weak interactions.

On the experimental side, very recent measurements of
single-- \cite{Ha01,Sa05,Ki02,Ok04}
and double--coincidence \cite{bhang,OutaVa,Kang05,Kim06} nucleon spectra 
from the non--mesonic hypernuclear decay were reported
---with accuracies largely improved with respect to the ones at disposal in
previous experiments \cite{Mo74,Sa91,Sz91,No95,No95a}--- in forms that suggest suitable
comparisons with theory. Some of these experiments have somehow managed to derive
values of $\Gamma_n/\Gamma_p$. They are discussed in the following paragraphs.

The $\Gamma_n/\Gamma_p$ ratio was obtained
by KEK--E307 \cite{Ha01,Sa05} for $^{12}_\Lambda$C, $^{28}_\Lambda$Si and $_\Lambda$Fe
hypernuclei from single--proton kinetic energy spectra measurements and by making use
of the intranuclear cascade code
of Ref.~\cite{Ra97} (based on the polarization propagator formalism of Ref.~\cite{Ra94})
to simulate the spectra of the nucleons emitted by the considered hypernuclei.
For $^{12}_\Lambda$C, $\Gamma_n/\Gamma_p=0.87\pm 0.23$ ($0.60^{+0.25}_{-0.23}$)
was obtained \cite{Sa05} by neglecting the two--nucleon induced decay mechanism (for
$\Gamma_2/(\Gamma_n+\Gamma_p)=0.35$, with $\Gamma_2\equiv \Gamma(\Lambda np \to np)$).
As the very same
authors of Refs.~\cite{Ha01,Sa05} noted in Ref.~\cite{Ki02}, these determinations of the
ratio may be affected by the fact that in the experiment the
neutron--induced decay width was estimated indirectly, from the proton measurement, using
the relation $\Gamma_n = \Gamma_{\rm T}-\Gamma_p-\Gamma_{\pi^-}-\Gamma_{\pi^0}$,
which neglects two--nucleon stimulated non--mesonic decays.
This method also required the measurement of the total decay width $\Gamma_{\rm T}$ as
well as the decay rates for the mesonic channels $\Lambda \to \pi^- p$ ($\Gamma_{\pi^-}$) and
$\Lambda \to \pi^0 n$ ($\Gamma_{\pi^0}$,
for which previous data from Ref.~\cite{Sa91} were used in the analysis of Ref.~\cite{Sa05}).
Moreover, the severe energy losses suffered by protons
inside the (thick) target and detector materials and the consequently
high kinetic energy threshold (about $40$ MeV) for proton detection in KEK--E307
did not permit an easy reconstruction of the proton spectrum emitted by the 
nucleus, which is essential for the indirect 
evaluation of $\Gamma_n$. Theoretical input about nucleon
rescattering in the residual nucleus \cite{Ra97} was indeed necessary to supply
to this problem. As a consequence, in Ref.~\cite{Ki02} the hypothesis was advanced
that $\Gamma_n$ ($\Gamma_p$) might be overestimated
(underestimated) in the analysis of Ref.~\cite{Sa05} because of an underestimation
in the number of emitted protons.

A controversial determination of the ratio from KEK--E369 data,
based on non-demonstrated, delicate hypotheses and theoretical
input (again from Ref.~\cite{Ra97}), was reported in
Ref.~\cite{Ki02}. In this experiment, direct measurements of
single--neutron kinetic energy spectra were performed (with a $10$
MeV threshold) for $^{12}_\Lambda$C and $^{89}_\Lambda$Y; once
analyzed together with the single--proton spectra of
Refs.~\cite{Ha01,Sa05}, a ratio $\Gamma_n/\Gamma_p=0.51\pm 0.15$
for $^{12}_\Lambda$C was derived by neglecting the two--nucleon
induced decay channel. We shall comment on the reliability of the extraction method
used for such a determination in Section~\ref{subsec:np}.

To overcome the difficulties of the discussed KEK experiments,
both single--neutron and single--proton energy spectra were measured
simultaneously by KEK--E462 for $^5_\Lambda$He and KEK--E508 for $^{12}_\Lambda$C
\cite{Ok04}. From these measurements,
the authors concluded that $\Gamma_n/\Gamma_p\simeq (N_n/N_p-1)/2\simeq 0.5$
for both $^5_\Lambda$He and $^{12}_\Lambda$C, $N_n$ ($N_p$) being the
total number of neutrons (protons) with kinetic energies $T_N$ above $60$ MeV. However, one
has to note that the previous approximate relation between $\Gamma_n/\Gamma_p$ and $N_n/N_p$
is only valid when FSI and two--nucleon induced decay effects can be neglected.
The predictions of Ref.~\cite{prc}
and the results presented in Sections \ref{single} and \ref{subsec:np}
prove that FSI are not negligible
even when a high detection threshold such as $T^{\rm th}_N=60$ MeV is used.

In the experiments KEK--E462 ($^5_\Lambda$He) and KEK--E508 ($^{12}_\Lambda$C),
nucleon--nucleon coincidence spectra were also measured
\cite{bhang,OutaVa,Kang05,Kim06}.
Quite clean angular and energy correlations between
neutron--neutron and neutron--proton emitted pairs
(i.e., back--to--back kinematics and $T_{N_1}+T_{N_2}\simeq 155$ MeV) were observed,
thus representing the first direct experimental evidence of the
existence of the two--body decays $\Lambda n\to nn$ and $\Lambda p\to np$.
The ratio, $N_{nn}/N_{np}$, between the numbers of emitted neutron--neutron and
neutron--proton pairs was measured to be around $0.5$
for both $^5_\Lambda$He and $^{12}_\Lambda$C after applying
the angular and energy restrictions: $\cos \theta_{NN}\leq -0.8$ and $T_N\geq 30$
MeV. The authors of Ref.~\cite{OutaVa,Kang05} concluded that, under these constraints,
$\Gamma_n/\Gamma_p\simeq N_{nn}/N_{np}\simeq 0.5$ on the basis of a supposed cancellation
of FSI and two--nucleon stimulated decays effects. 
In a very recent work
\cite{Kim06}, the result $\Gamma_n/\Gamma_p=0.51\pm 0.13\pm 0.04$ 
was deduced by the KEK collaboration 
for $^{12}_\Lambda$C after correcting for FSI effects by making use
of the number of detected proton--proton pairs in addition to 
measurements of $N_{nn}$ and $N_{np}$. 
A rather schematic method, not completely reliable in our opinion,  
quite in line with the one used in Ref.~\cite{Ki02}, was applied
to determine $\Gamma_n/\Gamma_p$. We shall discuss the effect of FSI and
two--nucleon induced decays in the extraction of $\Gamma_n/\Gamma_p$ from
measurements of $N_{nn}/N_{np}$ in Section \ref{subsec:nnnp}.

Despite this recent experimental progress, improved and/or independent measurements
are awaited for a really complete understanding of the $\Lambda N\to nN$ reaction
in nuclei. On this respect, the observation of the weak decay of
neutron and proton rich hypernuclei \cite{fi,hyphi} would also be source of
new information.

On the theoretical side, Refs.~\cite{prl,prc} presented an extensive
study of single-- and double--coincidence nucleon spectra
for the non--mesonic decay of $^5_\Lambda$He and $^{12}_\Lambda$C hypernuclei.
A one--meson--exchange (OME) model was used for the $\Lambda N\to nN$
transition in a finite nucleus framework. The two--nucleon induced decay channel
$\Lambda np \to nnp$ was taken into account via the polarization propagator method
in the local density approximation of Refs.~\cite{Ra94,Al00}.
The intranuclear cascade code of Ref.~\cite{Ra97}
was used to simulate the nucleon propagation inside the residual nucleus.
Comparison with KEK--E462 and KEK--E508 coincidence data
\cite{bhang,OutaVa,Kang05} lead to
the determination of $\Gamma_n/\Gamma_p$ values
around 0.3-0.4 for both $^5_\Lambda$He and $^{12}_\Lambda$C.
The relationship between $\Gamma_n/\Gamma_p$ and the observable
ratio between neutron--neutron and neutron--proton pairs, $N_{nn}/N_{np}$, was
established for the first time in Ref.~\cite{prc}.
It was shown that FSI and two--nucleon induced decays
significantly affect the extraction of $\Gamma_n/\Gamma_p$ from coincidence data
even when favorable energy and angular correlation
restrictions are imposed on the observed nucleon pairs.
Coincidence measurements of course make the determination of $\Gamma_n/\Gamma_p$
easier and cleaner, since FSI and two--nucleon induced decay effects are reduced
with respect to the ones emerging in analyses of single nucleon
observables such as $N_n/N_p$. Nevertheless, contrary to what claimed in
Refs.~\cite{bhang,OutaVa,Kang05}, they do not permit an exclusive identification of
the non--mesonic decay channels as neutron-- and proton--induced. Thus,
$\Gamma_n/\Gamma_p\neq N_{nn}/N_{np}$ (see figures 11 and 12 of Ref.~\cite{prc}
and the present discussion in Section \ref{subsec:nnnp}), indicating that
to determine $\Gamma_n/\Gamma_p$ one has to rely on experimental
as well as theoretical coincidence spectra.

With the purpose of making the extraction of $\Gamma_n/\Gamma_p$ less model dependent,
in the present paper we make use of an alternative framework for hypernuclear decay:
it consists of a nuclear matter formalism \cite{Ba03,Ba04}
extended to finite nuclei via the local
density approximation, with the same weak OME transition potential (containing
$\pi$, $\rho$, $K$, $K^*$, $\omega$ and $\eta$ exchange) of Ref.~\cite{Pa02},
in addition to the Monte Carlo intranuclear cascade code of Ref.~\cite{Ra97}.
At variance with previous analyses, in the present paper we follow a microscopic approach
for the two--nucleon induced decay, also including the channels $\Lambda nn \to nnn$ and
$\Lambda pp \to npp$ besides the standard mode $\Lambda np \to nnp$.
Most of the results shown are for the intermediate--mass
hypernucleus $^{12}_\Lambda$C, but we also present some results for a
heavier hypernucleus, $^{89}_\Lambda$Y. Both cases
can be well described within a local density approximation.

The paper is organized in the following way. The weak decay models employed
in the calculation are outlined in Section~\ref{weakdecay}. In Section~\ref{inc}
we give a few details of the intranuclear cascade simulation.
Numerical results for single and double coincidence
nucleon distributions are presented and compared with data in Section~\ref{results}.
The contribution of the two--nucleon induced decay channels is analyzed with
special regard. Finally, in Section~\ref{conclusion} we draw our conclusions.

%*********************************************************************
\section{Models}
%*********************************************************************
\label{models}
%*********************************************************************
\subsection{Weak decay}
%*********************************************************************
\label{weakdecay}
The weak decay transitions as well as the distributions of the weak decay nucleons
are obtained by means of a many--body
description of the $\Lambda$ self--energy in nuclear matter, based
on the polarization propagator method (PPM) originally proposed by
Oset and Salcedo \cite{Os85}.
The local density approximation (LDA) is employed to extend the calculation to finite nuclei.
This approach was previously established \cite{Ba03,Ba04} to evaluate
the non--mesonic decay widths of $^{12}_\Lambda$C.
There is a major value of this model which by itself adds
sufficient novelty to the present work with respect to previous analyses
\cite{prl,prc} of the nucleon spectra from hypernuclear non--mesonic decay.
We must indeed emphasize that the present approach is more
microscopic (i.e., less phenomenological) than the previous ones.
Indeed, in addition to the generally considered $\Lambda np \to nnp$ channel, also the
$\Lambda nn \to nnn$ and $\Lambda pp\to npp$ mechanisms are now evaluated.
The present approach is then
more in line with the functional approach ---in the framework of
the bosonic loop expansion--- used in Ref.~\cite{Al00b}, with the difference that
a clear separation into one-- and two--nucleon induced channels is now possible.

%*********************************************************************
\subsubsection{One--nucleon induced decay}
%*********************************************************************
\label{oneN}

\begin{figure}
\begin{center}
%\mbox{\epsfig{file=fig1.eps,width=.6\textwidth}}
    \includegraphics[width = .5\textwidth]{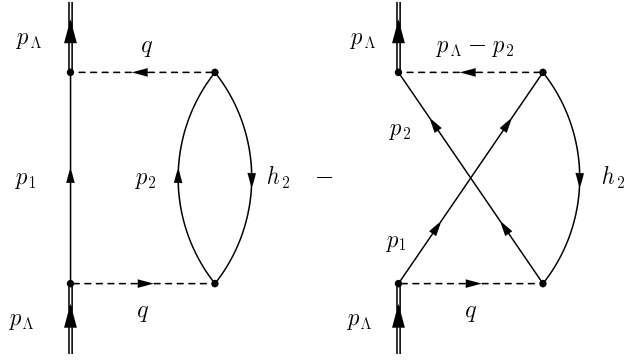}
\vskip 2mm
\caption{Direct and exchange 
$\Lambda$ self--energy diagrams corresponding to the one--nucleon induced decay channel
in nuclear matter.}
%\caption{Schematic representation of the transition amplitude for the
%one--nucleon induced channel in a nuclear medium. The $p_{i}$
%are the values of the energy--momentum of each particle.
%The dashed--line is the transition
%potential, $V^{\Lambda N\to nN}$, which
%carries an energy--momentum $q$ ($Q$) for the
%direct (exchange) contribution.}
\label{figme1}
\end{center}
\end{figure}

We start with a brief summary concerning the evaluation of
the one--nucleon ($1N$) induced decay widths $\Gamma_n$ and $\Gamma_p$.
Details can be found in Refs.~\cite{Ba03,Os85}. It is
convenient to work first with the partial decay width
$\Gamma_{t_{\Lambda} t_{N} \rightarrow t_{N'} t_{N''}}(p_{\Lambda},k_F)$,
where $p_{\Lambda}$ is the $\Lambda$ energy--momentum, $k_F$ is the Fermi
momentum of nuclear matter and the $t_i$'s represent the isospin projections 
of the baryons.
Using the standard Goldstone rules for diagrams in nuclear matter,
it is straightforward to write for the partial decay width:
\begin{eqnarray}
\label{gamdirexc}
\Gamma_{t_{\Lambda} t_{h2} \rightarrow
t_{p1} t_{p2}}(p_{\Lambda},k_F)  & =
& -2 \, {\rm Im} \int \frac{d^4 q}{(2 \pi)^4}
\int \frac{d^4  p_{\, 2}}{(2 \pi)^4} \; G_{\rm part}(p_{\Lambda}-q)\,
G_{\rm part}(p_{2})\, G_{\rm hole}(p_{2}-q)
\nonumber \\
&& \times \frac{1}{4} \sum \;
%\sum_{\tiny all \, spins}
\left(
\left|\bra{\gamma_{p1} \gamma_{p2}} V^{\Lambda N\to nN}(q)
\ket{\gamma_{\Lambda}
\gamma_{h2}}\right|^{2} \right.  \\
&& \left.
- \bra{\gamma_{p1} \gamma_{p2}} V^{\Lambda N\to nN}(q)
%(p_\Lambda-p_2)
\ket{\gamma_{\Lambda}
\gamma_{h2}}^{*}
\bra{\gamma_{p2} \gamma_{p1}} V^{\Lambda N\to nN}(p_\Lambda-p_2)
%(q)
\ket{\gamma_{\Lambda}
\gamma_{h2}}  \right), \nonumber
\end{eqnarray}
where $p_{i}$ ($h_{i}$) stands for the energy--momentum of particles (holes).
The meaning of each $p_{i}$ and $h_{i}$ is shown in Fig.~\ref{figme1}, where
we have drawn the $\Lambda$ self--energy diagrams from which the relevant transition amplitude
is obtained. From the energy--momentum conservation in each vertex we have
%$p_{1} = p_{\Lambda} - q$, $h_{2} = p_{2} - q$ and $Q = p_{\Lambda} - p_{2}$.
$p_{1} = p_{\Lambda} - q$ and $h_{2} = p_{2} - q$.
For simplicity,
$\gamma_{i}$ represents the spin and isospin projections
of particle~$i$. The one--meson--exchange
weak transition potential, $V^{\Lambda N\to nN}$,
takes into account the complete pseudoscalar and vector meson octets
($\pi,\eta,K,\rho,\omega,K^*$) through the parameterization of Ref.~\cite{Pa02}.
The summation in Eq.~(\ref{gamdirexc}) runs over all spins 
and isospins of the weak transition potential. 
Note that the first term in the
{\it r.h.s.}~of this equation is the usually called direct contribution, while
the second one is the exchange term. 
%As mentioned, direct and exchange transitions amplitudes are represented
%as the first and second diagrams in Fig.~\ref{figme1}, respectively.
%The square of the full amplitude lead to the decay width (\ref{gamdirexc}), where the square of
%each diagram is the direct contribution, while the interference term
%between the two diagrams is the exchange one.

The particle and hole propagators are, respectively:
\begin{eqnarray}
\label{nprop}
G_{\rm part}(p) & = & \frac{\theta(|\mbox{\boldmath $p$}| - k_F)}
{p_0 - E_N(\mbox{\boldmath $p$})- V_N + i
\varepsilon} \, ,  \\
%\mbox{~~~~and} \nonumber \\
G_{\rm hole}(h) & = & \frac{\theta(k_F - |\mbox{\boldmath $h$}| )}
{h_0 - E_N(\mbox{\boldmath $h$})- V_N - i
\varepsilon}\, , \nonumber
\end{eqnarray}
where $E_N(\mbox{\boldmath $p$})=m_N+\mbox{\boldmath $p$}^2/2m_N$ 
is the nucleon total free energy and $V_N$ the nucleon binding energy.

%The final decay width is obtained after integration over $p_{\Lambda}$ and $k_F$,
%and carrying out the summation over isospin, as follows,
By integrating Eq.~(\ref{gamdirexc}) over $\v{p}_{\Lambda}$ one obtains the $k_F$--dependent width:
\begin{equation}
%\label{decwpar2}
\Gamma_{t_{\Lambda} t_{h2} \rightarrow
t_{p1} t_{p2}}(k_F) = \int d \v{p}_{\Lambda} \,
\Gamma_{t_{\Lambda} t_{h2} \rightarrow
t_{p1} t_{p2}}(p_{\Lambda}, k_F) \;  |\psi_{\Lambda}(\v{p}_{\Lambda})|^2 \, , 
\end{equation}
where for the $\Lambda$ wave--function, $\psi_{\Lambda}(\v{p}_{\Lambda})$, we
take the $1s_{1/2}$ wave--function of a harmonic oscillator and
the $\Lambda$ energy, $(p_{\Lambda})_{0}$, is taken as
$(p_{\Lambda})_{0} = m_{\Lambda} + \v{p}_{\Lambda}^{2}/2m_{\Lambda} +
V_{\Lambda}$,
where $V_{\Lambda}$ is the experimental $\Lambda$ binding energy.
To evaluate the decay width for a particular nucleus, one uses
either an effective Fermi momentum or the LDA \cite{Os85}.
In the last case, $k_F$ is spatially dependent and the transition rate reads
\begin{equation}
%\label{decwpar3}
\Gamma_{t_{\Lambda} t_{h2} \rightarrow
t_{p1} t_{p2}} = \int d \v{r} \,
\Gamma_{t_{\Lambda} t_{h2} \rightarrow
t_{p1} t_{p2}}( k_F(r)) \;  |\widetilde{\psi}_{\Lambda}(\v{r})|^2 \, , 
\end{equation}
where $\widetilde{\psi}_{\Lambda}(\v{r})$ is the Fourier transform of
$\psi_{\Lambda}(\v{p}_{\Lambda})$.

Finally, by summing up over the isospin of baryons one obtains:
\begin{eqnarray}
\label{gam1np}
\Gamma_{n}
& \equiv & \Gamma_{\Lambda n \rightarrow n n}\, , \\
\Gamma_{p} & \equiv &
\Gamma_{\Lambda p \rightarrow n p} + \Gamma_{\Lambda p \rightarrow p n}\, , \nonumber
\end{eqnarray}
and the $1N$--induced decay width is:
\begin{equation}
%\label{gamma1}
\Gamma_1=\Gamma_n+\Gamma_p\, . \nonumber
\end{equation}
We have to stress that exchange terms are very important for an accurate evaluation
of $\Gamma_{1}$ \cite{Ba03}. The present work calculates exactly the exchange
terms in nuclear matter.
The Random Phase Approximation (RPA) corrections have not been taken into account,
because a precise evaluation of RPA--exchange terms
is a rather involved task, which goes beyond the scope of the
present contribution.

At this point we have to mention that
the models of Refs.~\cite{Pa02,Ba03,Ba04} have been recently improved. On the
one hand, a more realistic
$\Lambda$ wave--function, obtained in terms of the experimental hyperon binding energy,
has been considered in the LDA calculation of Refs.~\cite{Ba03,Ba04}, leading
to a reduction of about 25\% on the $1N$--induced decay rates,
$\Gamma_n$ and $\Gamma_p$, and of 30-40\% on the $2N$--induced ones,
$\Gamma_{nn}$, $\Gamma_{np}$ and $\Gamma_{pp}$, with respect to the original
results. On the other hand, a numerically more accurate evaluation of the
distorted final state wave--functions
has been implemented in the finite nucleus approach of Ref.~\cite{Pa02} and the
rates increase by about 15\%. In both cases the ratio
$\Gamma_n/\Gamma_p$ is essentially left unchanged. The updated predictions, used
in the present work, are given in Table~\ref{gammas}.
Fortunately, the spectra discussed in Refs.~\cite{prl,prc}, being normalized per non--mesonic
weak decay, are not affected by the corrections of Ref.~\cite{Pa02}.
%These spectra depends indeed on the ratio $\Gamma_n/\Gamma_p$, whose
%updated OMEa and OMEf values are only slightly changed with respect to the values,
%$0.29$ and $0.34$, of Ref.~\cite{Pa02}.
%Moreover, by definition, the weak decay model independent pair numbers of Table~V of
%\cite{Pa02}, used for the important purpose of
%extracting $\Gamma_n/\Gamma_p$ from measured spectra, remained unchanged.
Note that the non--mesonic rate $\Gamma_{NM}$ of the finite nucleus calculation
is underestimated by about 25\% because it lacks the contribution of the 
$2N$--stimulated mechanism. On the other hand, the LDA result is overestimated because
two--body final state interactions between the emitted nucleons were not
considered. As shown
in Ref.~\cite{Pa02}, the omission of this effect leads to rates larger by as much
as a factor of two. However, this factor disappears in the normalized
spectra discussed in the present work and used for the important purpose of
extracting $\Gamma_n/\Gamma_p$ from measured distributions.

\begin{table}
\begin{center}
\caption{Non--mesonic weak decay rates (in units of the free $\Lambda$ decay width)
predicted for $^{12}_\Lambda$C by the updated finite nucleus
approach (OMEa and OMEf calculations) of Ref.~\protect\cite{Pa02} and LDA model of
Refs.~\protect\cite{Ba03,Ba04}.}
\label{gammas}
\begin{tabular}{l c c c c c c c}
\mc {1}{c}{Ref.} &
\mc {1}{c}{$\Gamma_n$} &
\mc {1}{c}{$\Gamma_p$} &
\mc {1}{c}{$\Gamma_{nn}$} &
\mc {1}{c}{$\Gamma_{np}$} &
\mc {1}{c}{$\Gamma_{pp}$} &
\mc {1}{c}{$\Gamma_n/\Gamma_p$} &
\mc {1}{c}{$\Gamma_{\rm NM}$} \\ \hline
Finite Nucleus, OMEa      &  $0.190$ & $0.625$ &         &         &         &
$0.303$ & $0.815$  \\
Finite Nucleus, OMEf      &  $0.173$ & $0.484$ &         &         &         &
$0.356$ & $0.657$  \\
LDA &  $0.267$ & $0.936$ & $0.017$ & $0.238$ & $0.062$ & $0.285$  & $1.521$ \\ \hline
KEK--E508 \protect\cite{OutaVa} &  &  &  &  &  &  & $0.953\pm 0.032$ \\
KEK--E369 \protect\cite{Ki02}   &  &  &  &  &  & $0.51\pm0.15$ & \\
KEK--E307 \protect\cite{Sa05}   &  &  &  &  &  & $0.87\pm0.09\pm0.21$ & $0.828\pm 0.056\pm0.066$ \\
KEK \protect\cite{No95}         &  &  &  &  &  & $1.87\pm0.59^{+0.32}_{-1.00}$ &
$0.89\pm0.15\pm0.03$ \\
BNL \protect\cite{Sz91}         &  &  &  &  &  & $1.33^{+1.12}_{-0.81}$  & $1.14\pm0.20$
\end{tabular}
\end{center}
\end{table}

For a comparison with experiment, in Table~\ref{gammas} we also report data
for $\Gamma_{\rm NM}$ and $\Gamma_n/\Gamma_p$.
Only experimental determinations of $\Gamma_n/\Gamma_p$ obtained
from single--nucleon measurements are quoted. Large experimental errors affect
all but the most recent data, especially for $\Gamma_n/\Gamma_p$.
Almost all $\Gamma_n/\Gamma_p$ data appear to
strongly overestimate any theoretical prediction found in the literature.

%*********************************************************************
\subsubsection{Two--nucleon induced decay}
%*********************************************************************
\label{twoN}

%We give now the expression for $\Gamma_{2}$.
In the quasi--deuteron approximation, the $2N$--induced decay mode
turns out to be dominated by the process $\Lambda n p \rightarrow n n p$:
the meson emitted in the
$\Lambda$ vertex is mainly absorbed by an isoscalar neutron--proton
correlated pair. However, the isospin quantum
number also allows two others mechanisms, $\Lambda n n \rightarrow n n n$
and $\Lambda p p \rightarrow n p p$, which then contribute to
the decay rate $\Gamma_{2}$.
Here we briefly outline our main expressions needed for
the evaluation of this rate, which include all the just mentioned processes
within a nuclear matter framework.
A complete description of this decay channel can be found in Ref.~\cite{Ba04}.

At variance with $\Gamma_{1}$, we limit ourselves to direct terms.
This is because $\Gamma_{2}$ is originated from ground state
correlations (GSC). It was shown in Ref.~\cite{ba98} that,
in the case of electron scattering, the exchange terms for
the two particle--two hole polarization propagator
%$\Pi_{2p2h}$
can be neglected. In fact, in that work it was
concluded that exchange terms are not relevant in the graphs
originated from GSC.

\begin{figure}
\begin{center}
%\mbox{\epsfig{file=fig2.eps,width=.4\textwidth}}
    \includegraphics[width = .33\textwidth]{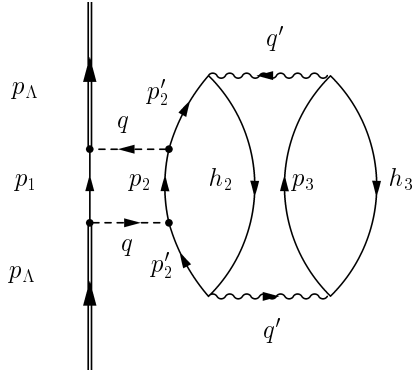}
\vskip 2mm
\caption{pp--part of the two particle--two hole contribution to the
$\Lambda$ self--energy in nuclear matter.}
\label{figme2}
\end{center}
\end{figure}

Furthermore, it is a good approximation \cite{Al84} to
consider only contributions where the two weak transition
potentials $V^{\Lambda N\to nN}$ (which again include the exchange of
$\pi$, $\rho$, $K$, $K^*$, $\eta$ and $\omega$ mesons\cite{Pa02}) of the
$\Lambda$ self--energy are attached to the same bubble.
%We follow the notation and ordering already displayed for $\Gamma_{1}$.
%As mentioned, the weak transition potential is attached to the same bubble.
There are thus three different self--energy contributions, which we denote by
$\rm pp$, $\rm ph$ and $\rm hh$. In the first one, the two $V^{\Lambda N\to nN}$ are
attached to the same particle (see Fig.~\ref{figme2}). In the
$\rm ph$ contribution one $V^{\Lambda N\to nN}$ is connected
to a particle and the other one to a hole. Finally,
the two potentials are attached
to the same hole for the $\rm hh$ part.
To illustrate the method, we present the pp contribution to $\Gamma_2$. 
From Goldstone rules, the $\rm pp$--partial decay width is:
\begin{eqnarray}
\label{gam2p2h}
\Gamma^{\rm pp}_{t_{\Lambda}t_{h2}t_{h3} \rightarrow
t_{p1}t_{p2} t_{p3}}(p_{\Lambda},k_F) & = &
 -2 \, {\rm Im} \, \int \frac{d^4 p_{2}}{(2 \pi)^4} \,
\int \frac{d^4  p_{3}}{(2 \pi)^4} \, \, \int \frac{d^4  q}{(2 \pi)^4}
\, \int \frac{d^4 q'}{(2 \pi)^4}  \\
&& \times \frac{1}{8} \sum G_{\rm part}(p_\Lambda-q) \; G_{\rm part}(p_{2}) \;
G^{2}_{\rm part}(p_{2}-q) \; G_{\rm part}(p_{3}) \nonumber \\
&&\times G_{\rm hole}(p_{2}-q+q') \; G_{\rm hole}(p_{3}-q') \nonumber \\
%&& \times \bra{\gamma_{h2} \gamma_{h3}}
%[V^{NN}(q')]^{\dag}
%\ket{\gamma_{p'2} \gamma_{p3}}
%\bra{\gamma_{\Lambda} \gamma_{p'2}}
%[V^{\Lambda N\to nN}(q)]^{\dag}
%\ket{\gamma_{p1} \gamma_{p2}} \nonumber \\
&& \times \left|\bra{\gamma_{p1} \gamma_{p2}}
V^{\Lambda N\to nN}(q)
\ket{\gamma_{\Lambda} \gamma_{p'2}}\right|^2
\left|\bra{\gamma_{p'2} \gamma_{p3}}
V^{NN}(q')
\ket{\gamma_{h2} \gamma_{h3}}\right|^2 . \nonumber
\end{eqnarray}
The meaning of each energy--momentum $p_{i}$ and $h_{i}$ is shown in Fig.~\ref{figme2}.
From energy--momentum conservation in each vertex we have:
$p_{1} = p_{\Lambda} - q$, $p'_{2} = p_{2} - q$,
$h_{2} = p_{2} - q + q'$ and $h_{3} = p_{3} - q'$.
Again, $\gamma_{i}$ stands for the spin and isospin projections of particle~$i$.
The nuclear residual interaction $V^{NN}$ is modeled by a
Bonn potential~\cite{Ma87}
in the framework of the parameterization presented in Ref.~\cite{Br96},
which contains the exchange of $\pi$, $\rho$, $\sigma$ and $\omega$ mesons
and neglects the small $\eta$ and $\delta$--mesons contributions.
The summation in
Eq.~(\ref{gam2p2h}) runs over all spins and 
%isospin of the weak
%transition potential and the nuclear residual interaction as well as
over $t_{{p'}_2}$.

After performing the LDA as described in the previous subsection, 
the summation over the isospin of baryons leads to:
\begin{eqnarray}
\label{gamnnnppp}
\Gamma^{\rm pp}_{n n}
& \equiv & \Gamma^{\rm pp}_{\Lambda n n \rightarrow n n n}\, , \\
\Gamma^{\rm pp}_{n p} & \equiv &
\Gamma^{\rm pp}_{\Lambda np \rightarrow nnp} + \Gamma^{\rm pp}_{\Lambda pn
\rightarrow nnp} +
\Gamma^{\rm pp}_{\Lambda np \rightarrow npn} + \Gamma^{\rm pp}_{\Lambda pn
\rightarrow npn} +
\Gamma^{\rm pp}_{\Lambda np \rightarrow pnn} + \Gamma^{\rm pp}_{\Lambda pn
\rightarrow pnn} \, , \nonumber \\
\Gamma^{\rm pp}_{pp} & \equiv &
\Gamma^{\rm pp}_{\Lambda pp \rightarrow npp} + \Gamma^{\rm pp}_{\Lambda pp
\rightarrow pnp} \, . \nonumber
\end{eqnarray}
Thus, once the $\rm ph$ and $\rm hh$ channels are included,
the $2N$--induced decay width is obtained as:
\begin{equation}
\Gamma_2=\Gamma_{nn}+\Gamma_{np}+\Gamma_{pp} \, , 
\end{equation}
where ($ij=nn$, $np$ or $pp$):
\begin{equation}
\Gamma_{ij}=\Gamma^{\rm pp}_{ij}+\Gamma^{\rm ph}_{ij}+\Gamma^{\rm hh}_{ij} \, . 
\end{equation}
Explicit expressions for all contributions can be found in Ref.~\cite{Ba04}.

The results obtained for the $2N$--stimulated decay widths
are given in Table~\ref{gammas}.
As expected, the dominant contribution to $\Gamma_2$ originates from the
$np$--induced decay, $\Lambda np \to nnp$. 
Nevertheless, the other $2N$--induced channels
cannot be neglected. The prediction $\Gamma_2/\Gamma_1=0.26$ is in agreement
with the phenomenological estimate $\Gamma_2/\Gamma_1=0.25$ of Refs.~\cite{prl,prc}.

%*********************************************************************
\subsection{Intranuclear cascade}
%*********************************************************************
\label{inc}

After the $\Lambda$ interacts with one (two) nucleons,  the two (three)
produced nucleons will interact with other nucleons in their way out of the
nucleus. This process, which will generate secondary nucleons, is accounted
for by the intranuclear cascade model described in Ref.~\cite{Ra97}. This model
considers a semiclassical
propagation of primary (i.e., weak decay) and secondary nucleons.
Nucleons move along classical, straight trajectories between collision points
and under the influence of a local (i.e., $R$--dependent)
mean potential, $V_N(R)=-k_{F_N}(R)^2/2m_N$,
where $k_{F_N}(R)=[3\pi^2 \rho_N(R)]^{1/3}$ is the local nucleon ($N=n$, $p$) Fermi momentum
corresponding to the nucleon density $\rho_N(R)$.
Propagating nucleons also collide with the nucleons of the medium according to free space
nucleon--nucleon cross sections properly corrected
to take into account the Pauli blocking and Fermi motion effects.
For more details of the code we refer to Ref.~\cite{Ra97}.

The description in terms of semiclassical nucleon propagation is
justified when the nucleon wavelength $\lambdabar$ is much smaller than the
average distance between nucleons, $d$, and the range of the nucleon--nucleon
interaction, $r_0$: $\lambdabar << r_0 \leq d$ \cite{Cu03}.
For a $30$ MeV kinetic energy nucleon,
$\lambdabar=0.8$ fm, which has to be compared with $d\simeq 2$ fm in $^{12}_\Lambda$C
and $r_0\simeq 1.4$ fm. The spectra for nucleon kinetic energies
$\lsim 30$ MeV could thus show unrealistic behaviors. For this reason, all the comparison
with data will intervene by introducing a kinetic energy cut of at least $30$ MeV.

%*********************************************************************
\section{Results}
%*********************************************************************
\label{results}
%*********************************************************************
\subsection{Single--nucleon spectra}
%*********************************************************************
\label{single}

\begin{figure}
\begin{center}
%\mbox{\epsfig{file=ldafn1.eps,width=.65\textwidth}}
    \includegraphics[width = .65\textwidth]{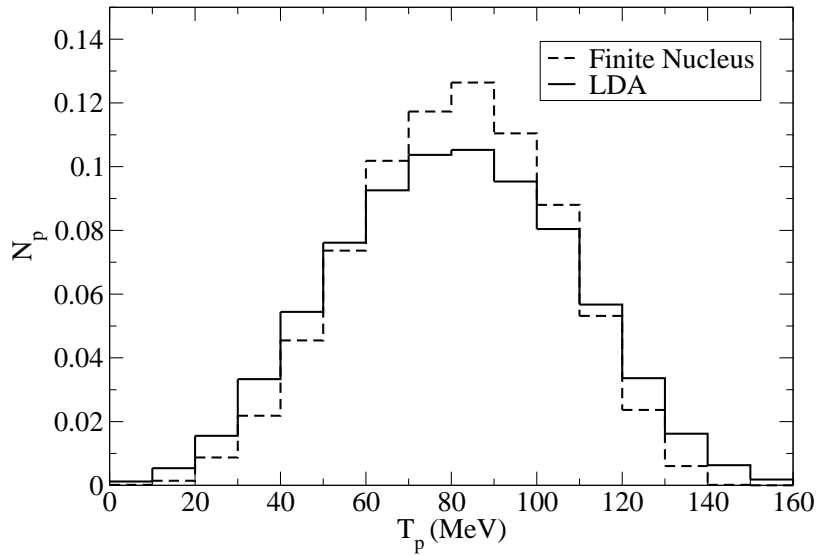}
\vskip 2mm
\caption{Kinetic energy spectra of primary protons from the $1N$--induced non--mesonic
weak decay of $^{12}_\Lambda$C. The continuous line refers to the present
LDA calculation, the dashed one to the finite nucleus evaluation
of Refs.~\protect\cite{prl,prc}. Both curves are normalized per $1N$--induced
decay assuming $\Gamma_n/\Gamma_p=(\Gamma_n/\Gamma_p)^{\rm LDA}=0.285$.}
\label{ldafn1}
\end{center}
\end{figure}

In Figure~\ref{ldafn1} a comparison is shown between the present
LDA calculation and the previous finite nucleus evaluation of
Refs.~\cite{prl,prc}. The number of primary protons emitted in the
$1N$--induced
non--mesonic weak decay of $^{12}_\Lambda$C is given as a function
of the proton kinetic energy. In order to make the comparison model independent,
the spectra are normalized per $1N$--induced
decay assuming $\Gamma_n/\Gamma_p=(\Gamma_n/\Gamma_p)^{\rm LDA}=0.285$, which is
equivalent to normalize per the corresponding proton--induced decay rate.
%(the total number of
%primary protons is $N^{\rm tot}_p=\Gamma_p/(\Gamma_n+\Gamma_p)=0.778$.
We observe that the Full Width a Half Maximum (FWHM) of the LDA distribution
is 15-20 MeV larger than the one of the finite nucleus spectrum. This is essentially due to
the more pronounced Fermi motion effects in the nuclear matter calculation.
Consistently with this expectation,
in the finite nucleus calculation of Refs.~\protect\cite{prl,prc} we found that the
primary nucleon distributions
for $^{12}_\Lambda$C were slightly broader than the ones for $^5_\Lambda$He.

\begin{figure}
\begin{center}
%\mbox{\epsfig{file=NpTp.eps,width=0.65\textwidth}}
    \includegraphics[width = .65\textwidth]{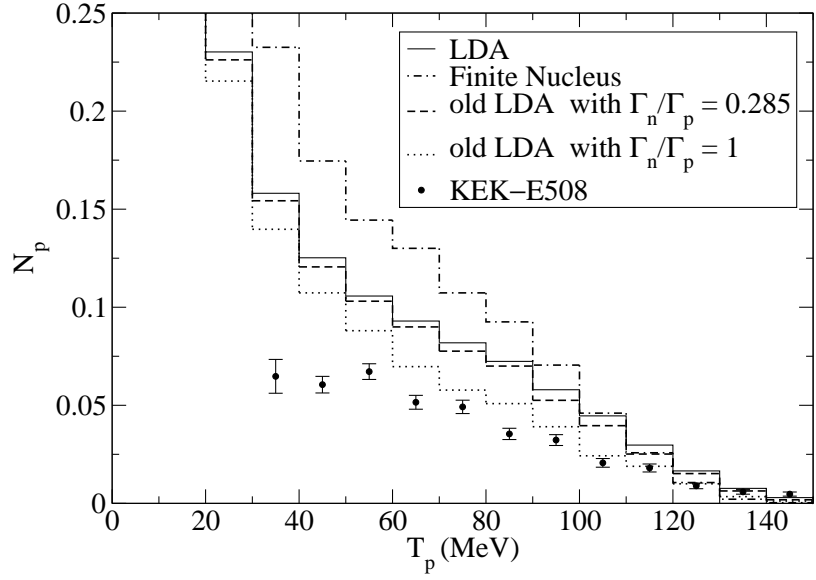}
\vskip 2mm
\caption{Single--proton kinetic energy spectra
for the non--mesonic weak decay of $^{12}_\Lambda$C after $2N$--induced
decays and FSI effects are included. Data are from KEK--E508 \protect\cite{Ok04}.
All results are normalized per non--mesonic weak decay.
See detailed explanation in the text.}
\label{ldafn2}
\end{center}
\end{figure}

In Figure~\ref{ldafn2} we show the single--proton kinetic energy spectra
for the non--mesonic weak decay of $^{12}_\Lambda$C once 
$2N$--induced decays and FSI effects are included. All spectra are normalized
per non--mesonic weak decay. The continuous line refers to the
present LDA result. The dashed line has been obtained with the LDA of
Ref.~\cite{Ra97}, where the weak transition potential was described
by a correlated pion exchange. The old
LDA spectrum corresponds to the same $\Gamma_n/\Gamma_p$ of the present LDA,
$\Gamma_n/\Gamma_p=(\Gamma_n/\Gamma_p)^{\rm LDA}=0.285$. Good
agreement is obtained between the two LDA calculations despite
the different weak decay models employed. The dot--dashed line
has been taken from
the finite nucleus evaluation of Refs.~\cite{prl,prc} fixing 
$\Gamma_n/\Gamma_p=(\Gamma_n/\Gamma_p)^{\rm LDA}=0.285$. Again, the differences
between LDA and finite nucleus estimates are mainly due to the different phase spaces
in the two cases. As it is apparent from Figure~\ref{ldafn2}, all
the theoretical spectra
are in strong disagreement with KEK--E508 data \cite{Ok04}. Not even a calculation
enforcing a large value of $\Gamma_n/\Gamma_p$ such as $1$ in the old LDA (dotted
curve) is able to reproduce the data. Surprisingly, such a LDA \cite{Ra97}
was able to reproduce the KEK--E307 single--proton data of Ref.~\cite{Sa05}
with values of $\Gamma_n/\Gamma_p$ smaller than 1,
%more specifically
%around 0.87 by neglecting the two--nucleon stimulated decay mechanism \cite{Sa05},
as reported in the Introduction.
This indicates that the two sets of data are inconsistent with each other. A clarification of
this discrepancy would be desirable.

\begin{figure}
\begin{center}
%\mbox{\epsfig{file=NnTn.eps,width=.65\textwidth}}
    \includegraphics[width = .65\textwidth]{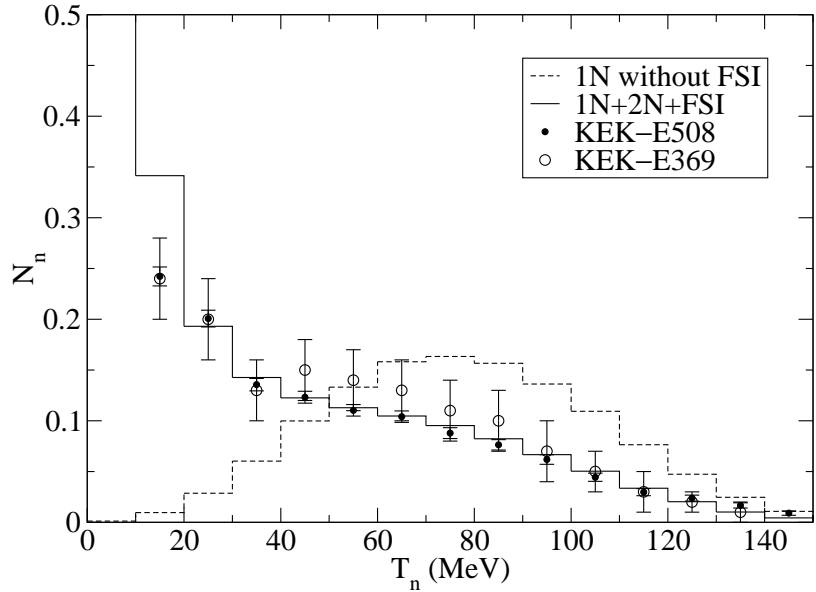}
\vskip 2mm
\caption{Single--neutron kinetic energy spectra for the non--mesonic
weak decay of $^{12}_\Lambda$C. The dashed line corresponds to the
distribution of primary neutrons, while the continuous line is obtained
once $2N$--induced decays and FSI effects are included.
Data are from KEK--E369 \protect\cite{Ki02} and
KEK--E508 \protect\cite{Ok04}. The spectrum of primary neutrons
(experimental data and the full theoretical result) is
(are) normalized per $1N$--induced (total) non--mesonic
weak decay.}
\label{lda2}
\end{center}
\end{figure}

In contrast, the single--neutron spectrum measured by KEK--E508 \cite{Ok04} is
compatible, within error bars, with the previous distribution measured by the
KEK--E369 experiment\cite{Ki02}, as can be seen in Fig.~\ref{lda2}. Here, results
of the present LDA calculation are also shown, both for the primary neutrons coming
from the $1N$--induced
mechanism (dashed line) and after including $2N$--induced decays and FSI effects (solid
line). The full result is in good agreement with both sets of data.

We note in passing that a model having a sufficiently large value of
$\Gamma_n/\Gamma_p$ to reproduce the KEK--E508 proton spectrum of
Fig.~\ref{ldafn2} would inevitably overestimate the neutron
spectra of KEK--E508 and KEK--E369 shown in Fig.~\ref{lda2}. 
%assuming a similar relative size of the two--nucleon induced
%contribution and FSI effects. 
We would also like to point out that the KEK--E508 proton
spectrum of Fig.~\ref{ldafn2}
have suffered a much stronger correction from energy losses in target and
detector materials than the neutron spectra of Fig.~\ref{lda2}.

We would like to comment now on a
disagreement between theory and experiment concerning the decay of $^5_\Lambda$He.
The calculation of Ref.~\cite{prc} predicted a quite pronounced peak at $T_n\simeq 75$ MeV
in the single--neutron energy spectra from $^5_\Lambda$He non--mesonic decay.
Even if a LDA formalism is not the best description for this light hypernucleus,
we have checked that a very similar peak is also produced by the present model.
In contrast, almost no peaking structure was found in the KEK--E462 experiment of Ref.~\cite{Ok04}.
%[Pregunta sobre la que hay que investigar: Are we sure that the neutron contamination from
%pion absorption was subtracted in the correct way from the
%observed neutron spectra? According to the theoretical evaluation of Ref.~\cite{Ok04},
%this background effect turned out to be very important in the whole neutron energy range.]

To gain some insight into this discrepancy, we put forward a simple
theoretical argument that shows that single--\emph{neutron} spectra are
expected to be less influenced
by FSI than single--\emph{proton} spectra. In this discussion we safely neglect
$2N$--stimulated decays. We start by noting that, for any value
of $\Gamma_n/\Gamma_p$, the number of primary neutrons is larger than that of
primary protons, i.e., 
$N^{\rm wd}_n/N^{\rm wd}_p=2\Gamma_n/\Gamma_p+1$ is always larger than $1$.
 Consequently,
due to $np\to np$ reactions, the proportion of secondary protons in the proton spectrum
$N_p$ (mainly at low kinetic energies) is larger than the proportion of secondary
neutrons in the neutron spectrum $N_n$. Note that $nn\to nn$ and $pp\to pp$ reactions,
occurring with almost identical cross sections, produce the same proportion of secondary
neutrons in $N_n$ and secondary protons in $N_p$, respectively.
For this reason, our maxima at $T_N\simeq 75$ MeV for $^5_\Lambda$He are more evident
for neutrons than for protons (compare Fig.~3 and 4 of Ref.~\cite{prc}). Due to stronger FSI,
such maxima completely disappear for $^{12}_\Lambda$C (compare Fig.~5 and 6 of Ref.~\cite{prc}),
but again protons are more affected by FSI than neutrons.

According to this discussion,
a similar shape of neutron and proton spectra ---as indicated by the experiment of
Ref.~\cite{Ok04} for $^{12}_\Lambda$C--- 
is only possible for $N^{\rm wd}_n\simeq N^{\rm wd}_p$ and thus for small
values of $\Gamma_n/\Gamma_p$. For $\Gamma_n/\Gamma_p=0$ we predict $N_n \simeq N_p$
 (see Fig.~\ref{singleratio}), but
KEK obtained $N_n/N_p\simeq 2$ for both $^5_\Lambda$He and $^{12}_\Lambda$C.
The similar neutron and proton KEK--E462 spectral shapes thus look surprising.
%One should note that the fact that neutron spectra turn out to be
%less affected by FSI than proton spectra is not due to a different propagation
%of neutrons and protons within the residual nucleus
%(the charge independence of the nucleon--nucleon interaction and the assumption
%of an isospin--symmetric propagation medium are reasonable approximations in these kind
%of discussions). It is indeed due to the different numbers of neutrons and protons produced
%in the $1N$--induced weak decay: $N^{\rm wd}_n=(2\Gamma_n +\Gamma_p)/\Gamma_1\geq
%N^{\rm wd}_p= \Gamma_p/\Gamma_1$.

Another point raised in the Conclusion of Ref.~\cite{Ok04} has to be commented.
There, Okada et al. claim that the observed
single--neutron spectrum from $^5_\Lambda$He ``indicates the importance of the multi--nucleon
induced process in the non--mesonic weak decay or/and a large FSI effect''.
Here we want to emphasize that only a very large and quite certainly unrealistic proportion
of $2N$--stimulated decays could eliminate the maximum that we found
at $T_n\simeq 75$ MeV for $^5_\Lambda$He decay. Besides,
the disappearance of this maximum would require a too strong amount of FSI.

\begin{figure}
\begin{center}
%\mbox{\epsfig{file=NnTn_ytrium.eps,width=.75\textwidth}}
    \includegraphics[width = .65\textwidth]{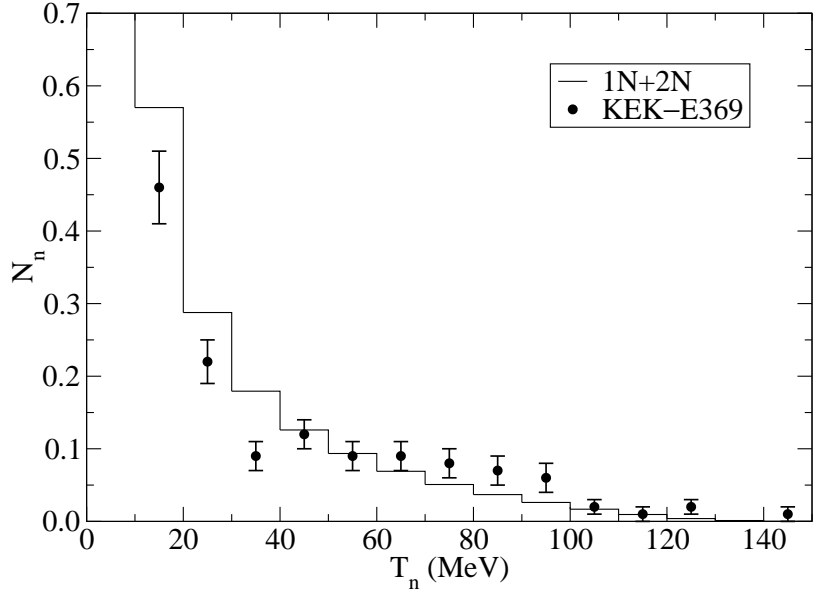}
\vskip 2mm
\caption{Single--neutron kinetic energy spectra for the non--mesonic
weak decay of $^{89}_\Lambda$Y. Data are from KEK--E369 \protect\cite{Ki02}.
All results are normalized per non--mesonic weak decay.}
\label{lda3}
\end{center}
\end{figure}

In Fig.~\ref{lda3} we show the single--neutron energy spectrum
obtained for the non--mesonic decay of $^{89}_\Lambda$Y. This result is
compared with the distribution obtained by the experiment KEK--E369 \cite{Ki02}.
The quite good agreement between theory
and data for a hypernucleus as heavy as $^{89}_\Lambda$Y
is an indicator of the reliability in using the intranuclear cascade code
of Ref.~\cite{Ra97} to simulate the nucleon FSI in hypernuclear decay.

%%%%%%%%%%%%%%%%%%%%%%%%%%%%%%%%%%%%%%%%%%%%%%%%%%%%%%%%%
\subsection{Ratio \mbox{\boldmath$N_n$/\boldmath$N_p$}}
%%%%%%%%%%%%%%%%%%%%%%%%%%%%%%%%%%%%%%%%%%%%%%%%%%%%%%%%%
\label{subsec:np}

We would like to compare now our results for the ratio between the number of
neutrons and the number of protons produced in the decay of $^{12}_\Lambda$C
with experimental observations.
Since our definitive aim is to determine $\Gamma_n/\Gamma_p$ by such a comparison,
it is convenient \cite{prc} to start by introducing the number of nucleons of type $N$
($N=n$ or $p$) produced in $n$--induced ($N^{\rm 1Bn}_N$),
$p$--induced ($N^{\rm 1Bp}_N$), $nn$--induced ($N^{\rm 2Bnn}_N$),
$np$--induced ($N^{\rm 2Bnp}_N$) and $pp$--induced ($N^{\rm 2Bpp}_N$) decays. By
normalizing these quantities per $n$--, $p$--, $nn$--,
$np$-- and $pp$--induced decay, respectively,
the total number of nucleons of the type $N$ normalized per non--mesonic
weak decay is given by:
\begin{equation}
\label{nratioeq}
N_N = \frac{N^{\rm 1Bn}_N \Gamma_n+N^{\rm 1Bp}_N \Gamma_p+
N^{\rm 2Bnn}_N \Gamma_{nn}+N^{\rm 2Bnp}_N \Gamma_{np}+N^{\rm 2Bpp}_N \Gamma_{pp}}
{\Gamma_n+\Gamma_p+\Gamma_{nn}+\Gamma_{np}+\Gamma_{pp}} .
\end{equation}
By definition, the nucleon numbers $N^{\rm 1Bn}_N$, $N^{\rm 1Bp}_N$, etc,
are independent of the model employed to describe the weak decay. 
Their values depend on the strong interaction part
of the problem, related to nucleon FSI, and on the framework (finite nucleus
or nuclear matter) used for treating the hypernuclear structure effects,
which produce different phase space factors.
The dependence on the weak decay model enters
Eq.~(\ref{nratioeq}) via the various partial decay widths.

In table~\ref{nratio} we report our results for the
weak decay model independent nucleon numbers in the case of
a kinetic energy threshold for nucleon detection of $T^{\rm th}_N=60$ MeV.
\begin{table}
\begin{center}
\caption{Predictions for the weak interaction model independent quantities
$N^{\rm 1Bn}_N$, $N^{\rm 1Bp}_N$, $N^{\rm 2Bnn}_N$, $N^{\rm 2Bnp}_N$, $N^{\rm 2Bpp}_N$
and for $N^{\rm 2B}_N$ of Eqs.~(\ref{nratioeq}) and
(\ref{nn/np})--(\ref{n_n}) for $^{12}_\Lambda$C
and nucleon kinetic energies $T_N\geq 60$ MeV.}
\label{nratio}
\begin{tabular}{c c c c c c}
\mc {1}{c}{$N^{\rm 1Bn}_n$} &
\mc {1}{c}{$N^{\rm 1Bp}_n$} &
\mc {1}{c}{$N^{\rm 2Bnn}_n$} &
\mc {1}{c}{$N^{\rm 2Bnp}_n$} &
\mc {1}{c}{$N^{\rm 2Bpp}_n$} &
\mc {1}{c}{$N^{\rm 2B}_n$} \\ \hline
   $0.91$  & $0.47$ & $0.67$ & $0.50$  & $0.17$ & $0.45$   \\ \\
   $N^{\rm 1Bn}_p$ & $N^{\rm 1Bp}_p$ & $N^{\rm 2Bnn}_p$ &
$N^{\rm 2Bnp}_p$ & $N^{\rm 2Bpp}_p$ & $N^{\rm 2B}_p$ \\ \hline
   $0.11$  & $0.54$ & $0.08$ & $0.21$  & $0.56$ & $0.27$
\end{tabular}
\end{center}
\end{table}
From Eq.~(\ref{nratioeq}) and our results of Tables~\ref{gammas} and
\ref{nratio} we then determine:
\begin{equation}
\label{singleg2}
\frac{N_n}{N_p}=1.33 \, .
\end{equation}
We have to note that this result is close to the ones obtained in the finite
nucleus calculation of Ref.~\cite{prc}, namely $N_n/N_p=1.38$ and 1.42, using the
OMEa and OMEf model, respectively.
If the $2N$--stimulated decay mode is neglected, the present calculation predicts
\begin{equation}
\left(\frac{N_n}{N_p}\right)^{\rm 1N}=1.28 \, .
\label{singlesing2}
\end{equation}
These results underestimate the value $N_n/N_p=2.00\pm 0.17$
obtained by KEK--E508 \cite{Ok04} for $T^{\rm th}_N=60$ MeV.
Such an occurrence is related to the disagreement between theory and experiment
for the single--proton spectra of Fig.~\ref{ldafn2}. Our results
of Eqs.~(\ref{singleg2}) and (\ref{singlesing2}) should also
be compared with the previous experimental determination,
$N_n/N_p=1.73\pm 0.22$, obtained 
from KEK--E369 and KEK--E307 data \cite{Ki02} and
for a detection threshold of about $40$ MeV.

It is interesting to mention that, on the contrary, for $^5_\Lambda$He,
KEK--E462 \cite{Ok04} measured $N_n/N_p=2.17\pm 0.22$, which fairly agrees
with the values obtained in Ref.~\cite{prc}, namely $N_n/N_p=1.78$ and 1.98 for
the OMEa and OMEf models, respectively.

In order to determine $\Gamma_n/\Gamma_p$ for $^{12}_\Lambda$C,
we now consider a weak decay model independent analysis of the mentioned
KEK--E508 data. From Eq.~(\ref{nratioeq}) written for neutrons
and protons one obtains:
\begin{equation}
\label{nn/np}
\frac{N_n}{N_p} = \frac{N^{\rm 1Bn}_n \displaystyle \frac{\Gamma_n}{\Gamma_p}+N^{\rm 1Bp}_n +
\frac{1}{\Gamma_1}\left(1+\frac{\Gamma_n}{\Gamma_p}\right)\left(
N^{\rm 2Bnn}_n \Gamma_{nn}+N^{\rm 2Bnp}_n \Gamma_{np}+N^{\rm 2Bpp}_n \Gamma_{pp}\right)}
{N^{\rm 1Bn}_p \displaystyle \frac{\Gamma_n}{\Gamma_p}+N^{\rm 1Bp}_p +
\frac{1}{\Gamma_1}\left(1+\frac{\Gamma_n}{\Gamma_p}\right)\left(
N^{\rm 2Bnn}_p \Gamma_{nn}+N^{\rm 2Bnp}_p \Gamma_{np}+N^{\rm 2Bpp}_p \Gamma_{pp}\right)} \, .
\end{equation}
The ratio $\Gamma_n/\Gamma_p$ is thus obtained in terms of our theoretical
values for $\Gamma_{nn}/\Gamma_1$, $\Gamma_{np}/\Gamma_1$ and $\Gamma_{pp}/\Gamma_1$
and from data for $N_n/N_p$ as:
\begin{equation}
\label{gngpn}
\frac{\Gamma_n}{\Gamma_p}=
\frac{\displaystyle N^{\rm 1Bp}_n+N^{\rm 2B}_n \frac{\Gamma_2}{\Gamma_1}
-\left(N^{\rm 1Bp}_p+N^{\rm 2B}_p \frac{\Gamma_2}{\Gamma_1}
\right)\frac{N_n}{N_p}}
{\displaystyle \left(N^{\rm 1Bn}_p+N^{\rm 2B}_p
\frac{\Gamma_2}{\Gamma_1} \right) \frac{N_n}{N_p}
-N^{\rm 1Bn}_n-N^{\rm 2B}_n \frac{\Gamma_2}{\Gamma_1}} ,
\end{equation}
where:
\begin{equation}
\label{n_n}
N^{\rm 2B}_N = \frac{N^{\rm 2Bnn}_N \Gamma_{nn}+N^{\rm 2Bnp}_N \Gamma_{np}+N^{\rm 2Bpp}_N \Gamma_{pp}}
{\Gamma_{nn}+\Gamma_{np}+\Gamma_{pp}} \, .
\end{equation}
Using our predictions of Tables~\ref{gammas} and \ref{nratio} together with
the datum of KEK--E508 \cite{Ok04}, $N_n/N_p=2.00\pm 0.17$, we obtain:
\begin{equation}
\label{eratio}
\frac{\Gamma_n}{\Gamma_p} = 0.95 \pm 0.21 \, .
\end{equation}
Neglecting the $2N$--stimulated channel the result is:
\begin{equation}
\left(\frac{\Gamma_n}{\Gamma_p}\right)^{\rm 1N} = 0.88 \pm 0.16 \, ,
\end{equation}
while enhancing arbitrarily the $2N$--induced rates by a factor of two we
obtain:
\begin{equation}
\label{eratio2}
\left(\frac{\Gamma_n}{\Gamma_p}\right)^{\Gamma_2\to 2\,\Gamma_2} = 1.02 \pm 0.27 \, .
\end{equation}

The sensitivity of the ratio $\Gamma_n/\Gamma_p$ to the values of the
$2N$--induced decay widths turns out to be moderate, especially
if the error bars originated by the datum adopted for $N_n/N_p$ are taken into account.
This is due not only to the negligible role of the $2N$--stimulated processes
in Eq.~(\ref{gngpn}) ($N^{\rm 1Bp}_{n(p)}$
and $N^{\rm 1Bn}_{n(p)}$ are larger than $N^{\rm 2B}_{n(p)} \Gamma_2/\Gamma_1$
since a quite high energy threshold is employed),
but also to the particular value of $N_n/N_p$ used in the analysis, which
causes a certain cancellation among the $2N$--stimulated contributions
in both the numerator and denominator of
Eq.~(\ref{gngpn}): $N^{\rm 2B}_n \simeq N^{\rm 2B}_p N_n/N_p$. 
Such a cancellation would be complete, thus leading to the extraction of the
same central value of $\Gamma_n/\Gamma_p$ for any value of $\Gamma_2$,
if a ratio $N_n/N_p=N^{\rm 2B}_n/N^{\rm 2B}_p=1.65$ were used in the analysis. 
%The use of a different $N_n/N_p$ would make the result for $\Gamma_n/\Gamma_p$
%more sensitive to the two--nucleon induced channel.

\begin{figure}
\begin{center}
%\mbox{\epsfig{file=singleratio.eps,width=.65\textwidth}}
    \includegraphics[width = .65\textwidth]{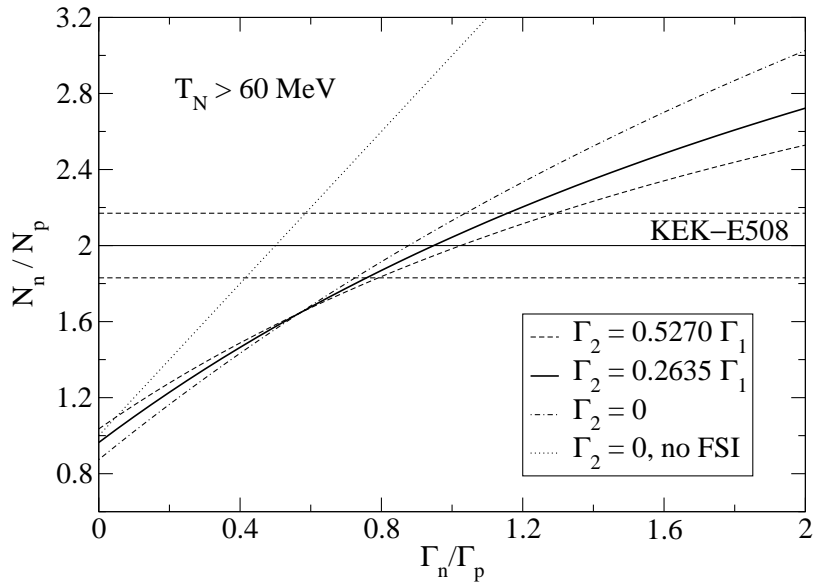}
\vskip 2mm
\caption{Dependence of the observable ratio $N_n/N_p$ on $\Gamma_n/\Gamma_p$
and $\Gamma_2/\Gamma_1$ for a nucleon energy threshold
of $60$ MeV. The horizontal lines show KEK--E508 data \protect\cite{Ok04}.
See text for further details.}
\label{singleratio}
\end{center}
\end{figure}

These occurrences can be understood from Figure~\ref{singleratio}, which shows the
relation between the observable ratio $N_n/N_p$ and $\Gamma_n/\Gamma_p$
for different choices of $\Gamma_2/\Gamma_1$. The dotted line corresponds
to the case in which $2N$--induced decays and FSI are neglected.
Once FSI are incorporated, quite different dependencies are obtained.
The dot--dashed line refers to the calculation in which $\Gamma_2$ is set
to $0$, the continuous line to the $\Gamma_2/\Gamma_1$ ratio predicted
by the present LDA model and the dashed line to the case in which
the size of $\Gamma_2$ is arbitrarily doubled.

Note in particular that the previously extracted $\Gamma_n/\Gamma_p$ values are 
different from the value 0.5 given in Ref.~\cite{Ok04} and
expected on the basis of the relation $\Gamma_n/\Gamma_p=(N_n/N_p-1)/2$, which
holds if $2N$--induced decays and FSI effects are ignored, i.e., if
$N^{\rm 1Bn}_n=2$, $N^{\rm 1Bp}_n=N^{\rm 1Bp}_p=1$, $N^{\rm 1Bn}_p=
N^{\rm 2Bnn}_{n(p)}=N^{\rm 2Bnp}_{n(p)}=N^{\rm 2Bpp}_{n(p)}=0$ in
Eqs.~(\ref{gngpn}) and (\ref{n_n}) (see the point where the experimental datum
intersects the dotted line in Fig.~\ref{singleratio}).
Contrary to the claim of Ref.~\cite{Ok04} and according to our results,
FSI turn out to be rather important even when one discusses the number ratio $N_n/N_p$
(for which part of the FSI effect is certainly canceled out)
and uses high kinetic thresholds such as $60$ MeV. On the contrary,
it is well manifest from Fig.~\ref{singleratio} that the 
$2N$--induced decay mechanism plays a relatively small role in the whole range
of reasonable $\Gamma_n/\Gamma_p$ values.

We also note that a na\"if estimation of FSI, such as that performed by
Kim et al.~\cite{Ki02} (which also neglect the $2N$--induced mechanism) 
taking a ratio $g/f=0.11$, where $f$ is the loss factor of each nucleon type and
$g$ the influx factor of one nucleon type converted from the other type,
gives $\Gamma_n/\Gamma_p=0.71\pm 0.14$
(see Eq.(4) of Ref.~\cite{Ki02}).
Although, due to error bars, this result is quite compatible
with the one obtained here with a more realistic treatment of FSI
and disregarding $2N$--stimulated decays,
$\Gamma_n/\Gamma_p=0.88 \pm 0.16$, clearly, the difference between the 
two central values demonstrates the following fact: that a detailed account of FSI
do not support the method used in Ref.~\cite{Ki02} to determine 
$\Gamma_n/\Gamma_p$.

We finalize this subsection by remarking that large ratios such as those of 
Eqs.(\ref{eratio})--(\ref{eratio2}) turn out to strongly overestimate the
value ($0.285$) predicted by the weak decay model employed in the present paper.
Actually, no calculation performed up to date reproduces these determinations,
which are more in agreement with the previous single--proton data of 
Refs.~\cite{Sa05,Sz91,No95}, as summarized in Table~\ref{gammas}. Only the
presence of an isoscalar, spin--independent central operator in the weak
transition potential can reproduce larger $\Gamma_n/\Gamma_p$ values, as can be
inferred from the effective treatment of Ref.~\cite{assum}.

\subsection{Double coincidence nucleon spectra}
%*********************************************************************
\label{double}

\begin{figure}
\vspace{1cm}
\begin{center}
%\mbox{\epsfig{file=nnangle.eps,width=0.65\textwidth}}
    \includegraphics[width = .65\textwidth]{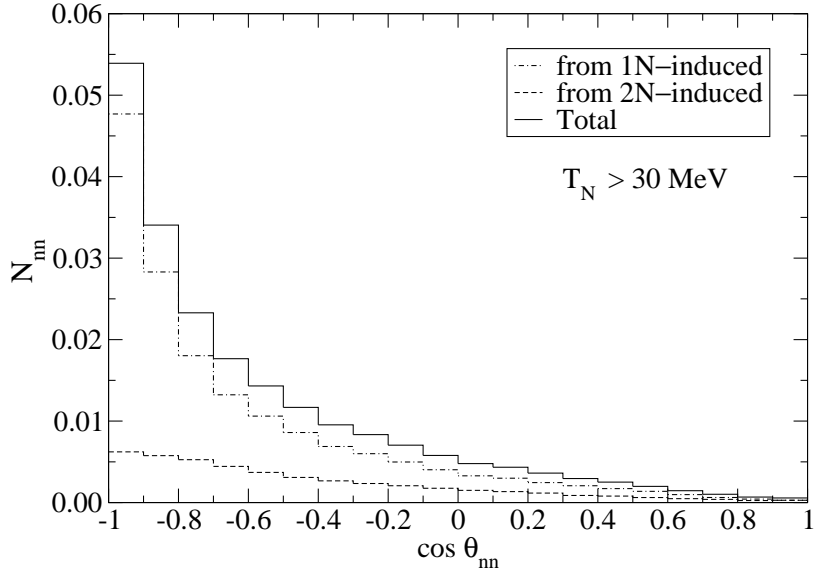}
\vskip 2mm
\caption{Opening angle distribution of $nn$ pairs normalized per
non--mesonic weak decay.}
\label{nnangle}
\end{center}
\end{figure}

\begin{figure}
\vspace{1cm}
\begin{center}
%\mbox{\epsfig{file=npangle.eps,width=0.65\textwidth}}
    \includegraphics[width = .65\textwidth]{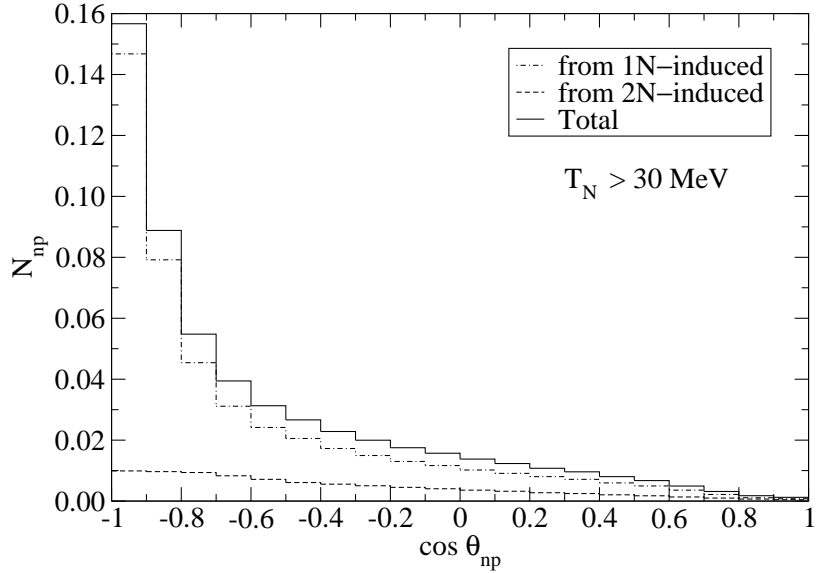}
\vskip 2mm
\caption{Opening angle distribution of $np$ pairs normalized per
non--mesonic weak decay.}
\label{npangle}
\end{center}
\end{figure}

Now we discuss the $NN$ coincidence spectra obtained from our model
for the decay of $^{12}_\Lambda$C. Figs.~\ref{nnangle} and \ref{npangle} show,
respectively, the distribution of $nn$ and $np$ pairs, as a function of the
cosine of the opening angle, where FSI effects have been incorporated and a
kinetic energy cut of 30 MeV has been applied.  We observe that the
distribution of pairs from the $1N$--induced processes (dash--dotted line) is more
back--to--back dominated than that coming from the  
$2N$--induced processes (dashed line). 
In any case, the $1N$--induced channel still provides the larger
contribution of pairs in the whole range of opening angles.
%In any case, the $1N$--induced process still represents the major component in the
%large opening angle region, the $2N$-induced mechanism only becoming more
%important in generating $nn$ ($np$) pairs for values of $\cos\theta$ larger
%than 0.4 (0.7).

For cos $\theta_{NN}\lsim -0.4$
the results of Figs.~\ref{nnangle} and \ref{npangle} reasonably reproduces the
ones of the previous finite nucleus calculation reported in
Fig.~5 of Ref.~\cite{prl} and Fig.~9 of Ref.~\cite{prc}.
On the contrary, for cos $\theta_{NN}\gsim -0.4$ a discrepancy 
%ascribable to the different phase space descriptions in the two approaches
is evident, the finite nucleus distributions being nearly flat in 
this region and the nuclear matter ones going monotonously to almost
vanishing values with cos $\theta_{NN}$. 
This is ascribable to the different phase space treatment 
in the two approaches. Indeed, the fact that the nuclear matter calculation
exhibits a more pronounced Fermi motion effect (see Fig.~\ref{ldafn1})
gives rise to a smaller number of outgoing nucleons with respect to the finite nucleus 
case (see Fig.~\ref{ldafn2}). This, in turn, is responsible for the fact that
the final nucleons of the finite nucleus calculation are more distorted by FSI than 
the nucleons in the nuclear matter approach. As a final result, 
the $NN$ opening angle distributions 
are more back--to--back correlated in the nuclear matter calculation.

Our distributions of Figs.~\ref{nnangle} and \ref{npangle} can also be compared with 
those obtained by KEK--E508 and shown in Figure~3 of the recent preprint by 
Kim et al.~\cite{Kim06}.
Due to the limited statistics of data, we concentrate on the angular region with 
cos $\theta_{NN}<-0.7$. In this region, experiment predicts $N_{nn}=0.083\pm0.014$
and $N_{np}=0.138\pm 0.014$, whereas our corresponding results are $N_{nn}=0.111$ and 
$N_{np}=0.300$, these numbers being normalized per non--mesonic weak decay. 
While there is decent agreement in the case of $N_{nn}$, for $N_{np}$
we overestimate the datum by a factor of about 2. The origin of such a 
discrepancy is likely the same as the one responsible for the disagreement on the
single--proton spectra of Fig.~\ref{ldafn2}. Another indication supporting this
hypothesis comes from comparing our result for the number
of proton--proton pairs for $T^{\rm th}_N=30$ MeV and 
cos~$\theta_{NN}<-0.7$, $N_{pp}=0.050$, with the experimental value $N_{pp}=0.005\pm 0.002$.
One cannot exclude the possibility that
the experiment systematically underestimated the number of protons emitted in
$^{12}_\Lambda$C decay, thus leading
to an underestimation of $N_p$, $N_{np}$ and $N_{pp}$ spectra. 
On the contrary, for the observables involving only neutron detection, $N_n$ and $N_{nn}$, 
theory and experiment are in reasonable accordance.

\begin{figure}
\begin{center}
%\mbox{\epsfig{file=nnkin.eps,width=.65\textwidth}}
    \includegraphics[width = .6\textwidth]{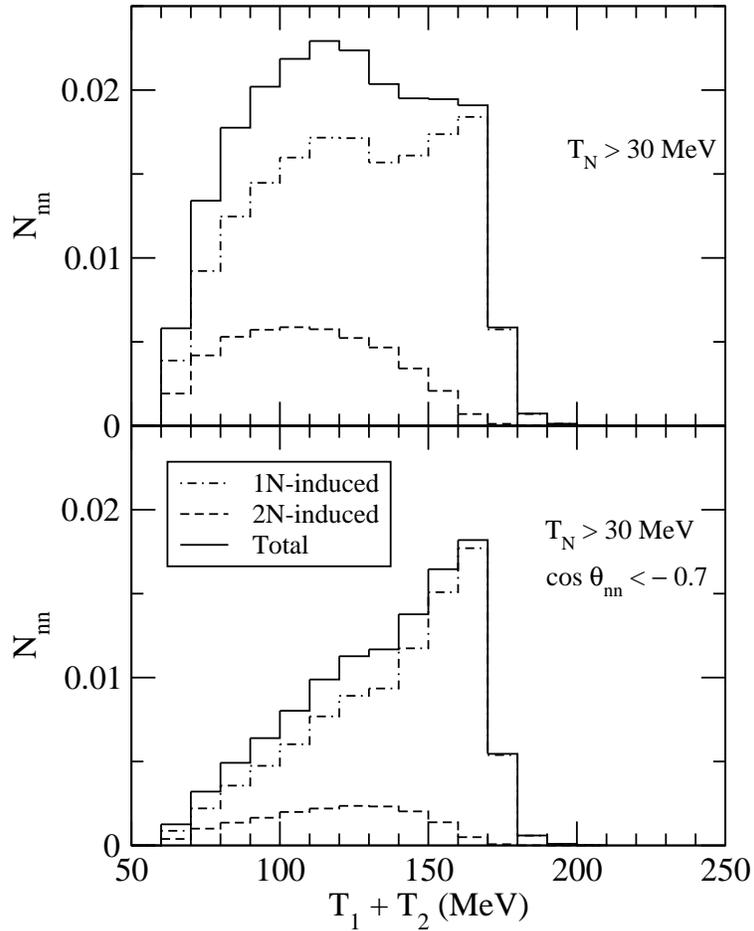}
\vskip 2mm
\caption{Distribution of the total kinetic energy of $nn$ pairs normalized 
per non--mesonic weak decay. The energy threshold is 30 MeV for each nucleon in
the pair. In the upper panel there is no angular restriction, while the condition
$\cos\, \theta_{nn} < -0.7$ has been imposed in the results of the lower panel.}
\label{nnkin}
\end{center}
\end{figure}

\begin{figure}
\begin{center}
%\mbox{\epsfig{file=npkin.eps,width=.65\textwidth}}
    \includegraphics[width = .6\textwidth]{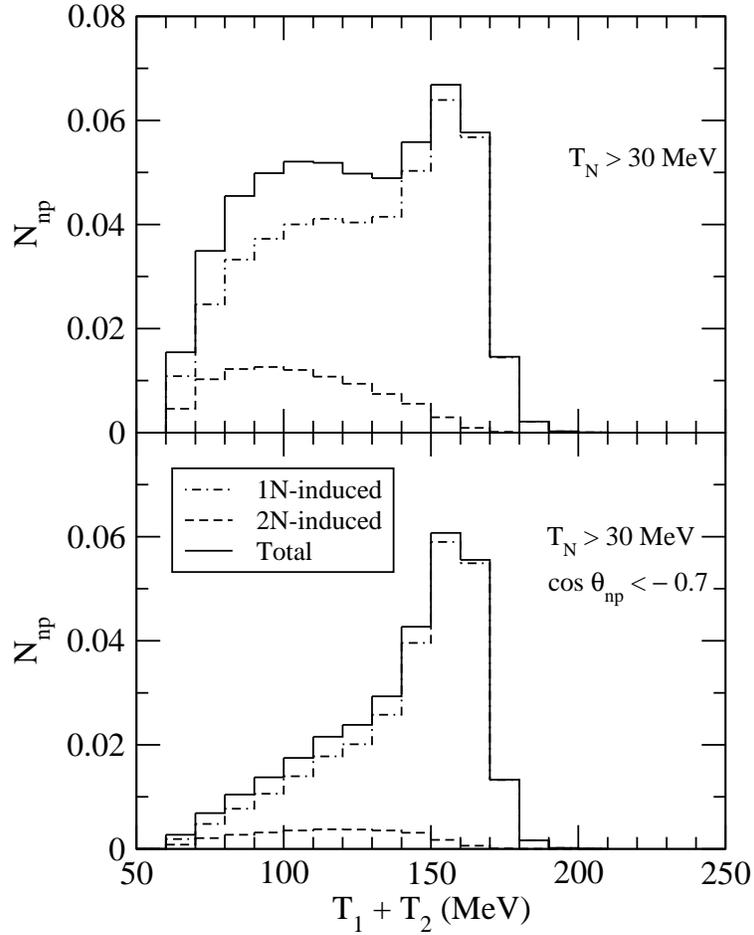}
\vskip 2mm
\caption{Distribution of the total kinetic energy of $np$ pairs normalized 
per non--mesonic weak decay. The energy threshold is 30 MeV for each nucleon in
the pair. In the upper panel there is no angular restriction, while the condition
$\cos\, \theta_{np} < -0.7$ has been imposed in the results of the lower panel.}
\label{npkin}
\end{center}
\end{figure}

The $nn$ and $np$ pair distribution as functions of the total kinetic energy of
the pair is shown in Figs.~\ref{nnkin} and \ref{npkin}, respectively, where the
energy of each nucleon is larger than a threshold kinetic energy of 30 MeV. The
upper panels show the distributions obtained without any cut in the opening
angle, while in the bottom panels only the back--to--back events are kept
by applying the restriction $\cos\, \theta_{NN} < -0.7$. 
We observe that the $2N$--induced events (dashed lines) in the upper panels 
scarcely contribute at the position of the
the primary peak of the $1N$--induced contribution (dot--dashed lines). 
Instead, they enhance the total 
distribution at a pair energy of around 100 MeV and generate a secondary peak
there, which might become even larger (see the case of the $nn$ pairs) than the
primary one at the $Q$--value of about 155 MeV. 
When the angular cut is applied, many of the events in
the low energy region are removed and the so--called back--to--back peak at 155
MeV stands out more clearly. However, there is still an important fraction of
the events (about 2/3 of $nn$ pairs and about 1/2 of $np$ pairs) that lie
outside this peak. This is in quantitative  agreement with 
the finite nucleus results of our previous works (see Fig.~3 of Ref.~\cite{prl}
and Fig.~10 of Ref.~\cite{prc}), the only qualitative difference being the
width of the back--to--back peak, which appears more smeared out in the
present work due to a more marked effect of Fermi motion.
In particular, note how the distributions from $1N$--induced decays
of Figs.~\ref{nnkin} and \ref{npkin}
extend above the $Q$--value of the non--mesonic decay.
%\begin{figure}
%\begin{center}
%%\mbox{\epsfig{file=nnkin08.eps,width=.95\textwidth}}
%    \includegraphics[width = .95\textwidth]{nnkin08.eps}
%\vskip 2mm
%\caption{Normalized to total (gamma1+gamma2) nmwd}
%\label{nnkin08}
%\end{center}
%\end{figure}

%\begin{figure}
%\begin{center}
%%\mbox{\epsfig{file=npkin08.eps,width=.95\textwidth}}
%    \includegraphics[width = .95\textwidth]{npkin08.eps}
%\vskip 2mm
%\caption{Normalized to total (gamma1+gamma2) nmwd}
%\label{npkin08}
%\end{center}
%\end{figure}

%%%%%%%%%%%%%%%%%%%%%%%%%%%%%%%%%%%%%%%%%%%%%%%%%%%%
\subsection{\mbox{\boldmath 2N}--induced strength}
%%%%%%%%%%%%%%%%%%%%%%%%%%%%%%%%%%%%%%%%%%%%%%%%%%%%

\label{2n}

In this subsection we compare the results of our microscopic model
for the $2N$--induced channel with those of the phenomenological
model of Ref.~\cite{Ra94}. We also analyze here the distribution of $NN$
pairs from the various sources of $2N$--induced strength.

\begin{figure}
\begin{center}
%\mbox{\epsfig{file=pdist2N.eps,width=.9\textwidth}}
    \includegraphics[width = .9\textwidth]{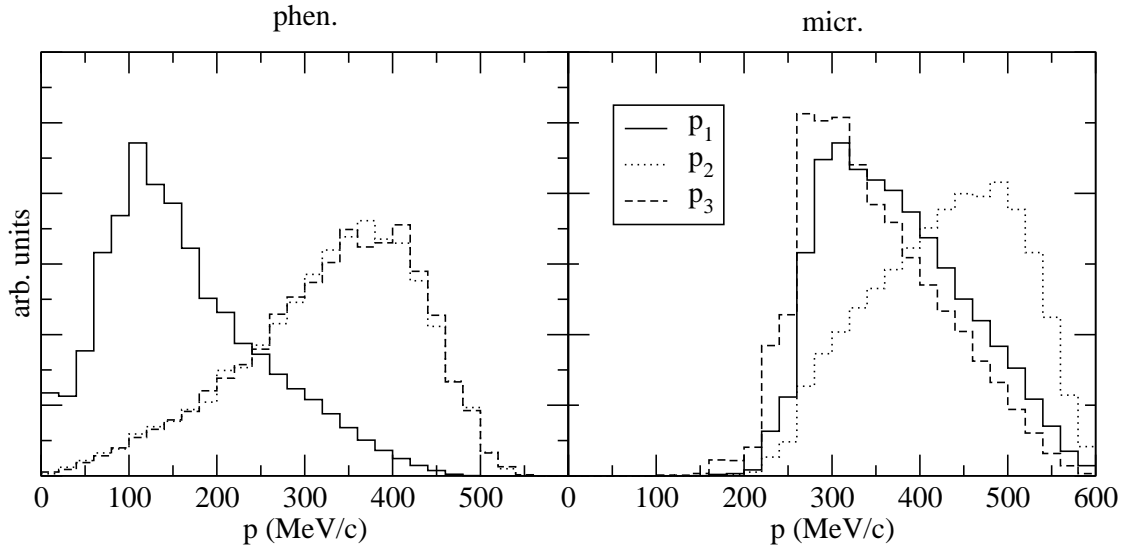}
\vskip 2mm \caption{Momentum distribution of each of the three
primary nucleons emitted in $2N$--induced processes.}
\label{fig:pdist2N}
\end{center}
\end{figure}

We start by commenting on the claim stated in Ref.~\cite{Ok04}
about the existence of two different types of $2N$--induced
decay. One would be a ``three--body reaction", in which the
available energy is equally shared among the three final nucleons, while
the other one should be due to the absorption of the pion emitted
at the weak vertex by a correlated two--nucleon pair. According to
Ref.~\cite{Ok04}, in our previous calculations \cite{prl,prc},
where the $2N$--induced strength was taken from the model of Ref.~\cite{Ra94},
we only included the latter. Here we want to clarify that there is
in fact only one type of $2N$--induced non--mesonic decay and 
that the particular kinematical properties of the emitted particles, prior
to FSI, are dictated by the dynamics of the $2N$--induced
decay mechanism. In the one--pion--exchange model
of Ref.~\cite{Ra94} the absorption by a correlated nucleon pair of
the virtual pion emitted at the $\Lambda$ vertex is driven by a
phenomenological two particle--two hole polarization
propagator $\Pi_{2p2h}(q^0,\mbox{\boldmath $q$})$. 
In the limit $(q^0,\mbox{\boldmath $q$}) \to (m_\pi,0)$ this propagator
describes the absorption of real pions in nuclear matter but we note that it was
conveniently extended, via a phase space correction, to account
for all possible $(q^0,\mbox{\boldmath $q$})$ values of the exchanged virtual
pion. Nevertheless, the pion at the $\Lambda$ vertex turns out to be
preferentially close to its mass shell and, consequently, the
nucleon emitted at the same vertex has very little kinetic energy
left, whereas the other two nucleons of the $2N$--induced channel are quite
energetic and come out in a back--to--back geometry. This is illustrated
in Fig.~\ref{fig:pdist2N}, that shows the distribution of momentum
values for each of the three nucleons emitted by the
$2N$--induced mechanism, prior to FSI effects. The left panel shows the
distribution obtained from the phenomenological model. It is
clearly seen that the nucleon emitted at the $\Lambda$ vertex, whose momentum 
is denoted by $p_1$, is
very slow while the other two are equally fast, having momenta on
average of about 400 MeV/c. In contrast, in the microscopic one--meson--exchange
model of Ref.~\cite{Ba04} used in the present work, the
nucleon emitted at the bubble that absorbs the exchanged meson
(see Fig.~\ref{figme2}) can reach high $p_2$ momentum values by
combining the momentum $q$ of the virtual meson plus the momentum
$p_2^\prime$ of the correlated nucleon that can be large due to
the short range nature of the $NN$ interaction. The momentum
$p_1$, carried by the nucleon emitted at the $\Lambda$ vertex, and
the momentum $p_3$, corresponding to the nucleon from the
spectator bubble, will acquire values characteristic of the range
of the interaction they come out from, which is, respectively, 
the weak and strong one--meson--exchange
models used in Ref.~\cite{Ba04}. As we observe in
the right panel of Fig.~\ref{fig:pdist2N}, the distributions of $p_1$
and $p_3$ momentum values turn out to be quite similar, peaking around 300
MeV/c, while that for $p_2$ peaks at higher momentum values close
to 500 MeV/c.

\begin{figure}
\begin{center}
%\mbox{\epsfig{file=ang2N.eps,width=.9\textwidth}}
    \includegraphics[width = .9\textwidth]{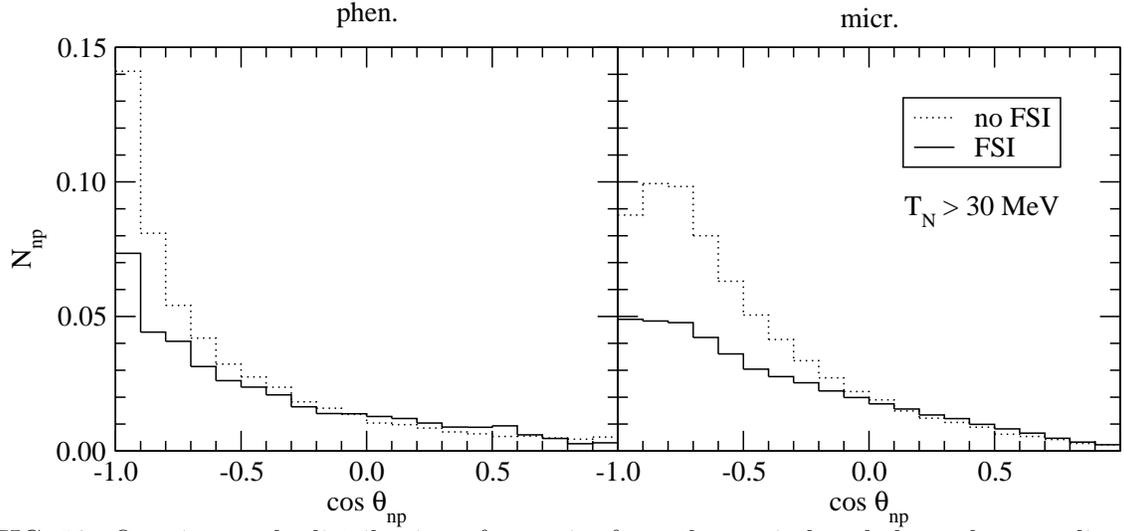}
\caption{Opening angle distribution of $np$ pairs from the $np$--induced
channel, normalized per $np$--induced decay.}
\label{fig:ang2N}
\end{center}
\end{figure}

\begin{figure}
\vspace{2.cm}
\begin{center}
%\mbox{\epsfig{file=doub2N.eps,width=.8\textwidth}}
    \includegraphics[width = 0.9\textwidth]{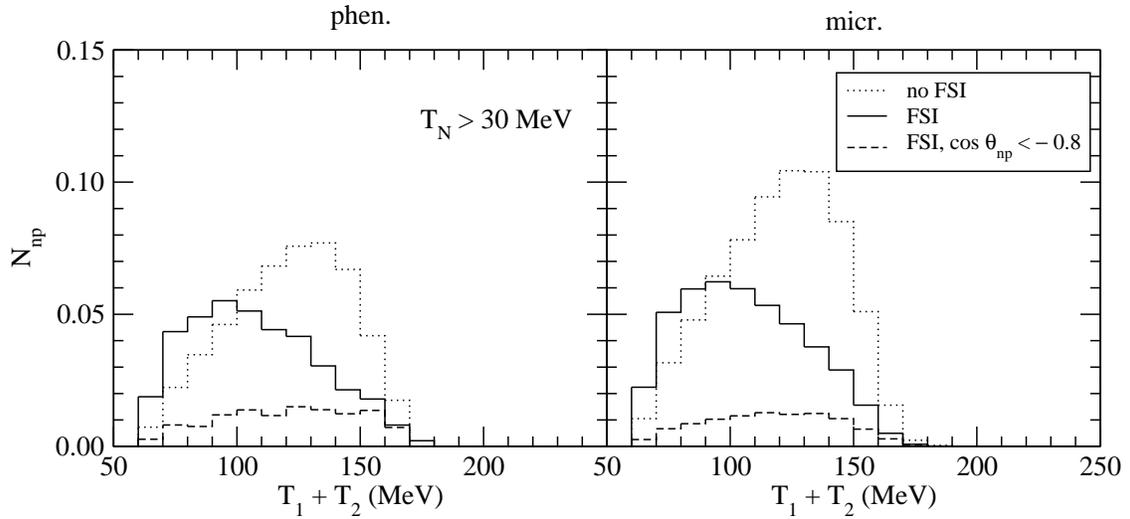}
\caption{Total kinetic energy distribution of $np$ pairs from the $np$--induced channel, 
normalized per $np$--induced decay.}
\label{fig:doub2N}
\end{center}
\end{figure}

We now compare the phenomenological and microscopic models after
the three primary nucleons emitted in the weak decay process are
allowed to undergo collisions with the other nucleons as they move
out of the nucleus. We restrict here to the $np$--induced decay mode,
which is the most important one in the microscopic approach and the only one considered by the
phenomenological model. The distribution of $np$ pairs, normalized
per its corresponding $np$--induced transition rate, is shown in
Fig.~\ref{fig:ang2N} as a function of the opening angle, where
the threshold kinetic energy is 30 MeV for
each nucleon of the pair. The dotted (solid) line
shows the results without (with) FSI effects. It is clear that, at
variance to the microscopical approach, the phenomenological model
produces a distinct back--to--back distribution which stands quite
clearly even when FSI effects are included. The distribution of pairs as a
function of the total kinetic energy and for a threshold $T^{\rm th}_N=30$ MeV
is displayed in Fig.~\ref{fig:doub2N}. We
observe that the amount of $N_{np}$ pairs per
$np$--induced decay event in the absence of FSI (area
under the dotted curves) is smaller in the phenomenological model, since the 
slow nucleon is always eliminated by the kinetic energy cut of 30 MeV. 
This situation is compensated when
the opening angle cut is also applied (dashed lines), since it removes more
events in the microscopic distribution, which is not so back--to--back dominated.
The conclusion is that, even if the kinematics of the primary nucleons look
quite different, at the end, once the effect of FSI is considered
and the energy and angular cuts are applied, both models produce similar neutron--proton
angular and energy spectra per $np$--induced decay event.
%, which are also distributed similarly over the total kinetic energy of the pair.

\begin{figure}
%\vspace{4.cm}
\begin{center}
%\mbox{\epsfig{file=nnnpangle.eps,width=.9\textwidth}}
    \includegraphics[width = 0.9\textwidth]{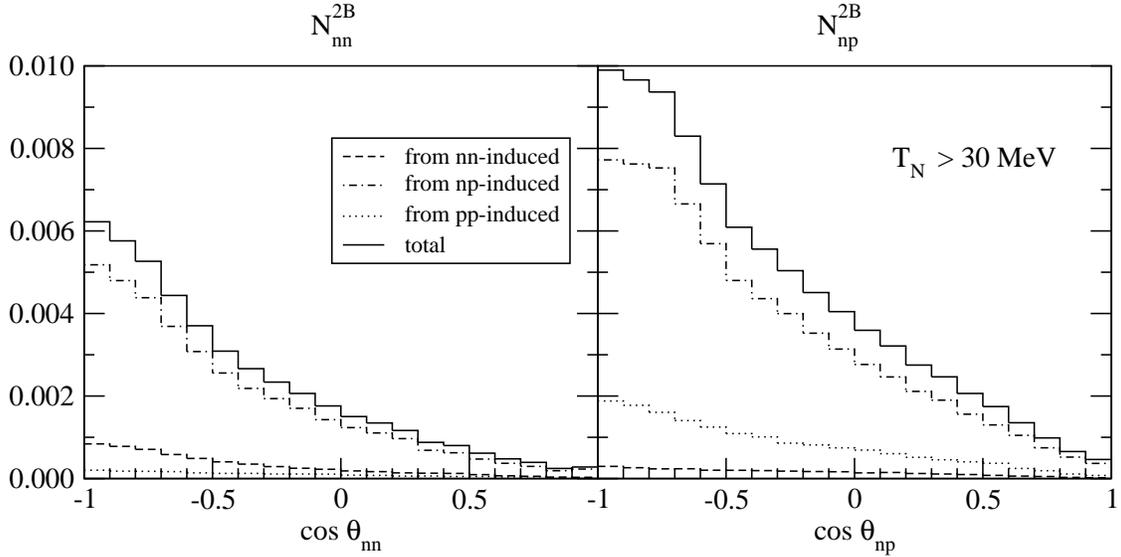}
\caption{Various $2N$--induced contributions ($2N=nn,np,pp$) to the
opening angle distribution of $nn$ (left) and $np$ pairs
(right). All results are normalized per total non--mesonic weak decay.}
\label{nnnpangle}
\end{center}
\end{figure}

We now discuss the distribution of the various contributions
($nn$--, $np$-- and $pp$--induced)
to the $2N$--induced non--mesonic decay in the microscopic model.
The opening angle distribution of $nn$ and $np$ pairs, again for
$T^{\rm th}_N=30$ MeV, is shown, 
respectively, on the left and right panels of Fig.~\ref{nnnpangle}. The angular
distribution of both $nn$ and $np$ pairs decreases smoothly with increasing
$\cos\, \theta_{NN}$. We observe that the
most important contribution is that of the $np$--induced decay, as we expected
from the values of the decay rates shown in Table~\ref{gammas}. We
note also that there is a small amount of $nn$ pairs from the
$pp$--induced process and $np$ pairs from the $nn$--induced one which would be
zero in the absence of FSI. 

\begin{figure}
\begin{center}
%\mbox{\epsfig{file=nnnpkin.eps,width=.9\textwidth}}
    \includegraphics[width = 0.9\textwidth]{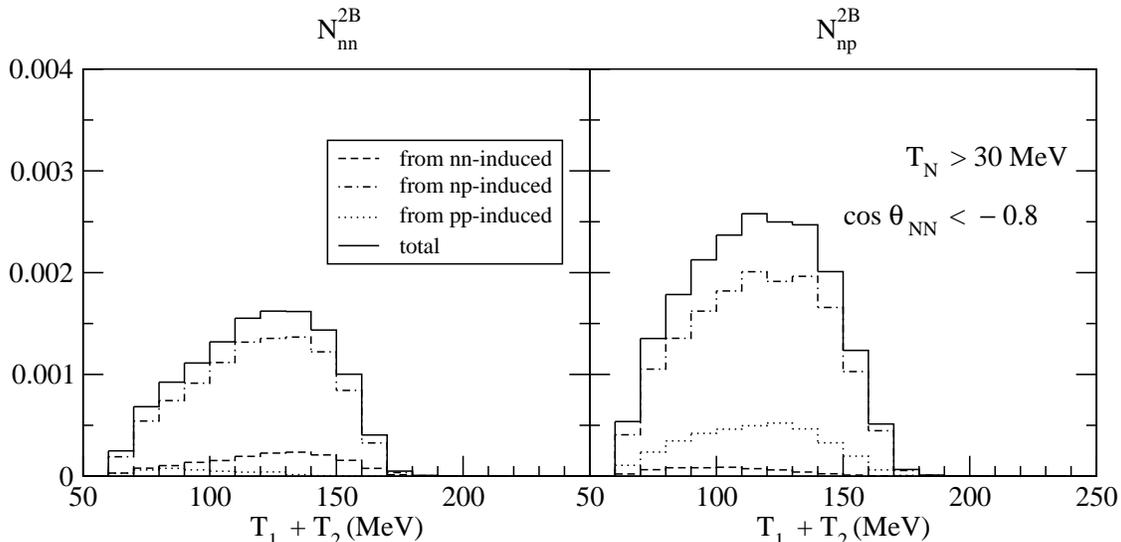}
\caption{Various $2N$--induced contributions ($2N=nn,np,pp$) to the
pair kinetic energy distribution of $nn$ (left) and $np$ pairs
(right). All results are normalized per total non--mesonic weak decay.}
\label{nnnpkin}
\end{center}
\end{figure}

The total pair energy distribution from the
$2N$--induced channels is shown in Fig.~\ref{nnnpkin}, where an angular cut of
$\cos\, \theta_{NN}<-0.8$ has been applied in addition to $T^{\rm th}_N=30$ MeV. 
The dominance of the $np$--induced decay
is obvious from figures \ref{nnnpangle}
and \ref{nnnpkin}, although after applying the usual 
kinetic and angular cut--offs the strength of 
some channels, as e.g. the $pp$--channel contributing to the number 
of $np$ pairs, can represent up to 20\% of the total. 

%%%%%%%%%%%%%%%%%%%%%%%%%%%%%%%%%%%%%%%%%%%%%%%%%%%%
\subsection{Ratio \mbox{\boldmath$N_{nn}/N_{np}$}}
%%%%%%%%%%%%%%%%%%%%%%%%%%%%%%%%%%%%%%%%%%%%%%%%%%%%%
\label{subsec:nnnp}

%In this section we obtain the ratio $\Gamma_n/\Gamma_p$ from the number of $nn$
%and $np$ pairs. 
We now discuss a comparison of our predictions for the ratio between the number
of $nn$ and $np$ pairs for $^{12}_\Lambda$C with experimental
determinations by KEK--E508 \cite{bhang,OutaVa,Kang05,Kim06}.
As in previous papers \cite{prl,prc}, we introduce the numbers
of $NN$ pairs ($NN=nn$, $np$ or $pp$) coming from one--nucleon induced 
($N_{NN}^{\rm 1Bn}$ and $N_{NN}^{\rm 1Bp}$) and two--nucleon induced 
($N_{NN}^{\rm 2Bnn}$, $N_{NN}^{\rm 2Bnp}$ and 
$N_{NN}^{\rm 2Bpp}$) processes, each one of them being
normalized per the rate of the corresponding process. These weak decay model 
independent quantities are
given in Table~\ref{univlda} for nucleon kinetic energies $T_N\geq 30$ MeV
and two angular regions. The total number of $NN$ pairs emitted
per non--mesonic weak decay event can be built from:
\begin{eqnarray}
\label{nratioNN}
N_{NN} &=& \frac{N^{\rm 1Bn}_{NN} \Gamma_n+N^{\rm 1Bp}_{NN} \Gamma_p+
N^{\rm 2Bnn}_{NN} \Gamma_{nn}+N^{\rm 2Bnp}_{NN} \Gamma_{np}+N^{\rm 2Bpp}_{NN} \Gamma_{pp}}
{\Gamma_n+\Gamma_p+\Gamma_{nn}+\Gamma_{np}+\Gamma_{pp}} \\
&=& \frac{N^{\rm 1Bn}_{NN} \Gamma_n+N^{\rm 1Bp}_{NN} \Gamma_p+
N^{\rm 2B}_{NN} \Gamma_2}{\Gamma_n+\Gamma_p+\Gamma_2} \, , \nonumber
\end{eqnarray}
where:
\begin{equation}
\label{NN2B}
N^{\rm 2B}_{NN} = \frac{N^{\rm 2Bnn}_{NN} \Gamma_{nn}+N^{\rm 2Bnp}_{NN} \Gamma_{np}+
N^{\rm 2Bpp}_{NN} \Gamma_{pp}} {\Gamma_{nn}+\Gamma_{np}+\Gamma_{pp}} 
\end{equation}
evidently depends on the weak model employed.
\begin{table}
\begin{center}
\caption{Predictions for the weak interaction model independent
quantities $N^{\rm 1Bn}_{NN}$, $N^{\rm 1Bp}_{NN}$,
$N^{\rm 2Bnn}_{NN}$, $N^{\rm 2Bnp}_{NN}$, $N^{\rm 2Bpp}_{NN}$ and 
for $N^{\rm 2B}_{NN}$ of Eqs.~(\ref{nratioNN}) and (\ref{NN2B}) 
for $^{12}_\Lambda$C, integrated over all opening angles and for nucleon energies
$T_N\geq 30$ MeV. The numbers in parentheses
correspond to the angular region with cos~$\theta_{NN}\leq -0.8$.}
\label{univlda}
\begin{tabular}{c c c c c c}
\mc {1}{c}{$N^{\rm 1Bn}_{nn}$} &
\mc {1}{c}{$N^{\rm 1Bp}_{nn}$} &
\mc {1}{c}{$N^{\rm 2Bnn}_{nn}$} &
\mc {1}{c}{$N^{\rm 2Bnp}_{nn}$} &
\mc {1}{c}{$N^{\rm 2Bpp}_{nn}$} &
\mc {1}{c}{$N^{\rm 2B}_{nn}$} \\ \hline
 $0.55$ ($0.31$) & $0.11$ ($0.03$) & $0.53$ ($0.15$) & $0.24$ ($0.06$)
& $0.05$ ($0.01$) & $0.22$ ($0.06$) \\ \\
 $N^{\rm 1Bn}_{np}$ & $N^{\rm 1Bp}_{np}$ &
 $N^{\rm 2Bnn}_{np}$ & $N^{\rm 2Bnp}_{np}$ & $N^{\rm 2Bpp}_{np}$ & $N^{\rm 2B}_{np}$ \\ \hline
 $0.34$ ($0.09$) & $0.64$ ($0.34$) & $0.27$ ($0.05$) & $0.45$ ($0.10$)
& $0.40$ ($0.09$) & $0.43$ ($0.09$) \\ \\
 $N^{\rm 1Bn}_{pp}$ & $N^{\rm 1Bp}_{pp}$ &
 $N^{\rm 2Bnn}_{pp}$ & $N^{\rm 2Bnp}_{pp}$ & $N^{\rm 2Bpp}_{pp}$ & $N^{\rm 2B}_{pp}$ \\ \hline
 $0.04$ ($0.01$) & $0.15$ ($0.05$) & $0.02$ ($0.004$) & $0.08$ ($0.02$)
& $0.30$ ($0.08$) & $0.12$ ($0.03$)
\end{tabular}
\end{center}
\end{table}

Before proceeding with the discussion on the ratio $N_{nn}/N_{np}$,
we want to compare the results of Table~\ref{univlda} with the ones obtained
within the finite nucleus framework of Refs.\cite{prl,prc} and reported 
%(once updated...?) 
in Table~\ref{univ}. A very good agreement is noticeable for the $1N$--induced
contributions. That is expected and depends on the fact that 1)
the considered numbers are independent of the weak decay model one uses 
and 2) the FSI
effects are modeled with the same intranuclear cascade code in both evaluations.
Possible differences have to be ascribed to the different frameworks 
(finite nucleus vs nuclear matter) used for describing hypernuclear structure effects.
The disagreement existing for some $2N$--induced contributions, whose significance
are anyhow relatively low for the total numbers $N_{NN}$
due to the smallness of the corresponding decay rates, 
is due to the different decay channels
%included in the two calculations 
and (especially) phase space factors involved in the two determinations 
(for a comparison of the phase spaces, we refer to the discussion of Fig.~\ref{fig:pdist2N}).
\begin{table}
\begin{center}
\caption{Predictions of the finite nucleus calculation of Ref.~\protect\cite{prc}
for the weak interaction model independent quantities
$\protect N^{\rm 1Bn}_{NN}$, $\protect N^{\rm 1Bp}_{NN}$ and for
$\protect N^{\rm 2B}_{NN}$ (integrated over all angles and for nucleon energies
$\protect T_N\geq 30$ MeV) for $\protect ^{12}_\Lambda$C.
The numbers in parentheses correspond to the angular region with
$\protect \cos \theta_{NN}\leq -0.8$.}
\label{univ}
\begin{tabular}{c c c}
\mc {1}{c}{$N^{\rm 1Bn}_{nn}$} &
\mc {1}{c}{$N^{\rm 1Bp}_{nn}$} &
\mc {1}{c}{$N^{\rm 2B}_{nn}$} \\ \hline
   $0.57$ ($0.31$) & $0.11$ ($0.03$) & $0.30$ ($0.12$)   \\ \\
    $N^{\rm 1Bn}_{np}$ & $N^{\rm 1Bp}_{np}$ & $N^{\rm 2B}_{np}$ \\ \hline
  $0.34$ ($0.09$) & $0.68$ ($0.32$) & $0.39$ ($0.10$) \\ \\
    $N^{\rm 1Bn}_{pp}$ & $N^{\rm 1Bp}_{pp}$ & $N^{\rm 2B}_{pp}$ \\ \hline
  $0.04$ ($0.01$) & $0.17$ ($0.05$) & $0.05$ ($0.01$)
\end{tabular}
\end{center}
\end{table}

We now come back to the $N_{nn}/N_{np}$ ratio.
From Eq.~(\ref{nratioNN}) and our results of Tables~\ref{gammas} and \ref{univlda} we obtain:
\begin{equation}
\label{NN1}
\frac{N_{nn}}{N_{np}}=0.36
\end{equation}
for the case with $\cos \theta_{NN}\leq 0.8$. This is in good agreement with the 
KEK--E508 datum $0.40\pm 0.10$ \cite{bhang}.
%of table~\ref{dataNN}.
%We have to note that this result is also close to the ones obtained in Ref.~\cite{prc}
%with the OMEa (0.39) and OMEf (0.43) models.
The comparison of the result (\ref{NN1}) with those obtained with the
calculations of Ref.~\cite{prc} (from Tables~\ref{gammas} and \ref{univ}),
namely $N_{nn}/N_{np}= 0.43$ and 0.47, using the
OMEa and OMEf models, respectively,
show the differences in the partial decay rates and phase spaces
predicted by the different models.
If the $2N$--stimulated decay mode is neglected, the present calculation predicts:
\begin{equation}
\label{NN2}
\left(\frac{N_{nn}}{N_{np}}\right)^{\rm 1N}=0.34 \, ,
\end{equation}
thus emphasizing a relatively small effect of the $2N$--induced channels on correlation
observables appropriately chosen.
%By comparing results (\ref{NN1}) and (\ref{NN2}) with 
%(\ref{singleg2}) and (\ref{singlesing2}) one also sees that, as expected, the effect
%of $2N$--induced decays is more important for single--nucleon
%than for double--coincidence observables.
The above result (\ref{NN2}) should be compared with the na\"if estimation
$N_{nn}/N_{np}=\Gamma_n/\Gamma_p=0.285$ obtained by neglecting both FSI and the $2N$--induced
decay channel: contrary to what occurs to single--nucleon spectra, i.e., to $N_n/N_p$,
the effect of FSI in the present case
of coincidence observables turns out to be not too marked.
The reliability of the above na\"if equality improves by using
higher thresholds $T^{\rm th}_N$ and more restrictive back--to--back constraints. 
%Calculate for 70 or 90 MeV?. Figure~\ref{doubleratio}
We can therefore conclude that, for the cases of experimental interest 
such as those considered here and in Section \ref{subsec:np},
the equation $N_{nn}/N_{np}=\Gamma_n/\Gamma_p$ turns out to have a 
wider range of approximate validity than the assumption $N_n/N_p=(2\Gamma_n+\Gamma_p)/\Gamma_p$
sometimes used in experimental analyses, 
%which indeed reveals to be almost inapplicable.
whose utilization to estimate $\Gamma_n/\Gamma_p$ must be avoided.
%(on this question, compare also Figs.\ref{singleratio} and \ref{doubleratio}).

We now consider a weak decay model independent analysis of the $N_{nn}/N_{np}$ data of
Table~\ref{dataNN} to determine $\Gamma_n/\Gamma_p$.
%(dependence only on the phase space factor, which is different in LDA and FN..).
These data have been obtained by KEK for $^5_\Lambda$He \cite{Kang05} and 
$^{12}_\Lambda$C \cite{bhang,Kim06} with a threshold $T^{\rm th}_N=30$ MeV
and various angular restrictions.
%\footnote{Una nota para nosotros:
%For $\cos \theta_{NN}\leq -0.7$, $\Gamma_n/\Gamma_p=0.56\pm 0.12$ is determined for
%$^{12}_\Lambda$C (Outa NPA paper in \cite{OutaVa}) after evaluating FSI using
%$pp$ pairs (not Bhang's $\beta$--method) and without two--nucleon induced mechanism.
%No entiendo como lo han sacado!}
The ratio $\Gamma_n/\Gamma_p$ is determined by using theoretical
values for $\Gamma_{nn}/\Gamma_1$, $\Gamma_{np}/\Gamma_1$ and $\Gamma_{pp}/\Gamma_1$
and data for $N_{nn}/N_{np}$ from:
\begin{equation}
\label{gngpnn}
\frac{\Gamma_n}{\Gamma_p}=
\frac{\displaystyle N^{\rm 1Bp}_{nn}+N^{\rm 2B}_{nn} \frac{\Gamma_2}{\Gamma_1}
-\left(N^{\rm 1Bp}_{np}+N^{\rm 2B}_{np} \frac{\Gamma_2}{\Gamma_1} \right)\frac{N_{nn}}{N_{np}}}
{\displaystyle \left(N^{\rm 1Bn}_{np}+N^{\rm 2B}_{np}
\frac{\Gamma_2}{\Gamma_1} \right) \frac{N_{nn}}{N_{np}}
-N^{\rm 1Bn}_{nn}-N^{\rm 2B}_{nn} \frac{\Gamma_2}{\Gamma_1}} \, ,
\end{equation}
where $N^{\rm 2B}_{NN}$ is given by Eq.~(\ref{NN2B}).
From our predictions of Tables~\ref{gammas} and \ref{univlda} and the $^{12}_\Lambda$C 
datum of KEK--E508 
derived with the restrictions $T_N\geq 30$ MeV and cos $\theta_{NN}\leq -0.8$ we obtain:
\begin{equation}
\label{con}
\frac{\Gamma_n}{\Gamma_p} = 0.34 \pm 0.15 .
\end{equation}
Besides, neglecting the $2N$--stimulated channel:
\begin{equation}
\label{sin}
\left(\frac{\Gamma_n}{\Gamma_p}\right)^{\rm 1N} = 0.37 \pm 0.14 ,
\end{equation}
while enhancing arbitrarily the $2N$--induced rates by a factor of two:
\begin{equation}
%\label{}
\left(\frac{\Gamma_n}{\Gamma_p}\right)^{\Gamma_2\to 2\,\Gamma_2} = 0.32 \pm 0.16 .
\end{equation}
The ratio (\ref{con}) [(\ref{sin})] must be compared with
$\Gamma_n/\Gamma_p=0.27 \pm 0.14$ [$\Gamma_n/\Gamma_p=0.36 \pm 0.14$]
obtained from the results of Ref.~\cite{prc} (listed in Table~\ref{univ})
together with a value of $\Gamma_2/\Gamma_1=0.25$ used there. 
While the ratios are very similar when $\Gamma_2=0$,
a certain difference is obtained for those including $2N$--stimulated decay effects.
This is quite obvious and
signals the different models used in the two works for the $2N$--induced channels.

\begin{figure}
\begin{center}
%\mbox{\epsfig{file=nnnpratio.eps,width=0.65\textwidth}}
    \includegraphics[width = .65\textwidth]{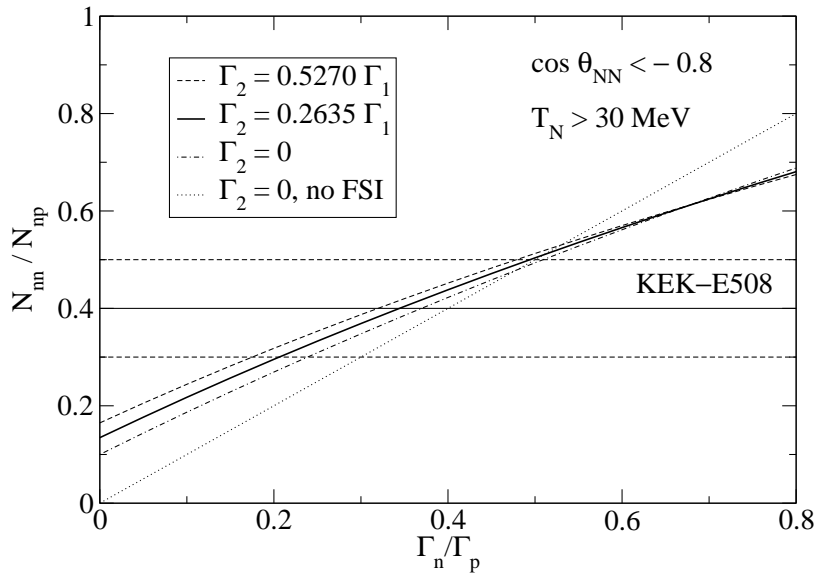}
\vskip 2mm
\caption{Dependence of the observable ratio $N_n/N_p$ on $\Gamma_n/\Gamma_p$
and $\Gamma_2/\Gamma_1$ for a nucleon energy threshold
of $30$ MeV and cos $\theta_{NN}\leq -0.8$.
The horizontal lines show KEK--E508 data \protect\cite{bhang}.
See text for further details.}
\label{doubleratio}
\end{center}
\end{figure}

In Figure~\ref{doubleratio} we show the
relation between $N_{nn}/N_{np}$ and $\Gamma_n/\Gamma_p$
for different choices of $\Gamma_2/\Gamma_1$ and for
$T^{\rm th}_N=30$ MeV and $\cos \theta_{NN}\leq -0.8$.
The dotted line corresponds
to the case in which $2N$--induced decays and FSI are neglected.
Different behaviours are obtained
once FSI are incorporated.
The dot--dashed line refers to the calculation in which $\Gamma_2$ is set
to $0$, the continuous line to the $\Gamma_2/\Gamma_1$ ratio predicted
by the present LDA model and the dashed line to the case in which
the size of $\Gamma_2$ is arbitrarily doubled. The comparison of
Fig.~\ref{doubleratio} with Fig.~\ref{singleratio} clearly illustrates the fact
that FSI affect much more the extraction of $\Gamma_n/\Gamma_p$ from  
$N_{n}/N_{p}$ than from $N_{nn}/N_{np}$.

Table~\ref{dataNN} summarizes the values of $\Gamma_n/\Gamma_p$ derived in
our analysis of KEK data for $N_{nn}/N_{np}$ in $^{12}_\Lambda$C. 
Also reported is the result for $^5_\Lambda$He of the finite nucleus calculation
of Ref.~\cite{prc}. In all cases the values reported for $\Gamma_n/\Gamma_p$
have been obtained by including the $2N$--stimulated contributions as predicted
by the two different approaches. As a signal of the limited statistics of the
data employed, 
the central value for $\Gamma_n/\Gamma_p$ in $^{12}_\Lambda$C shows a certain undesired 
dependence on the opening angle region. Nevertheless, the error bars
are so big that values in the interval 0.4-0.5 are common to all determinations. 
The final result can thus be given by the weighted average 
of the three partial results:
% lead to the ``best" value:
\begin{equation}
\left(\frac{\Gamma_n}{\Gamma_p}\right)^{\rm best}=0.43\pm 0.10 .
\end{equation}
When $2N$--induced decay channels are neglected, a similar analysis of data
supplies a very close ``best" value:
\begin{equation}
\left(\frac{\Gamma_n}{\Gamma_p}\right)^{\rm 1N\,\, best}=0.46\pm 0.09 .
\end{equation}
\begin{table}
\begin{center}
\caption{$^5_\Lambda$He \protect\cite{Kang05} and $^{12}_\Lambda$C \protect\cite{bhang,Kim06}
KEK data for the ratio $N_{nn}/N_{np}$
corresponding to a nucleon threshold $T^{\rm th}_N=30$ MeV and different
opening angle regions. The corresponding
determinations of $\Gamma_n/\Gamma_p$
obtained in the present work and from the results of Ref.~\protect\cite{prc}
are also reported.}
\label{dataNN}
\begin{tabular}{c c c c c c c}
\mc {1}{c}{} & &
\mc {2}{c}{$^5_\Lambda$He} & &
\mc {2}{c}{$^{12}_\Lambda$C} \\
\hline
Angular range &
& $N_{nn}/N_{np}$ 
& $\Gamma_n/\Gamma_p$ \cite{prc} &
& $N_{nn}/N_{np}$ &
$\Gamma_n/\Gamma_p$ [this work] \\
\hline
$\cos \theta_{NN}\leq -0.9$ & &  &  & & $0.45\pm 0.12$ & $0.43 \pm 0.17$  \\
$\cos \theta_{NN}\leq -0.8$ & &  $0.45\pm 0.11$ & $0.27 \pm 0.11$  & & $0.40\pm 0.10$ 
& $0.34 \pm 0.15$  \\
$\cos \theta_{NN}\leq -0.7$ & & &  & & $0.60\pm 0.12$ & $0.66 \pm 0.24$ 
\end{tabular}
\end{center}
\end{table}

We want to note that for the case with $\cos \theta_{NN}\leq -0.7$, Kim et al.~\cite{Kim06}
have recently extracted a ratio $\Gamma_n/\Gamma_p=0.51\pm 0.13\pm 0.04$ by employing 
the number of detected proton--proton pairs in addition to the measurement of
$N_{nn}$ and $N_{np}$. A schematic method for accounting for FSI effects,
similar to the one employed
in Ref.~\cite{Ki02} for single--nucleon observables, 
was adopted for this purpose. In addition, 
$2N$--stimulated decay channels were neglected. The discrepancy between this
KEK determination and our detailed prediction of Table~\ref{dataNN} for
$\cos\theta < -0.7$ 
(compare especially the central values) demonstrate that 
analyses of data such as those of Refs.~\cite{Ki02,Kim06} show certain limitations.

%...Play with different $T^{\rm th}_N$ in order to extract $\Gamma_n/\Gamma_p$
%and $\Gamma_2/\Gamma_1$...

%%%%%%%%%%%%%%%%%%%%%%%%%%%%%%%
\section{Conclusions}
%%%%%%%%%%%%%%%%%%%%%%%%%%%%%%%
\label{conclusion}

We have presented a study of single-- and double--coincidence nucleon spectra
for the non--mesonic weak decay of $\Lambda$--hypernuclei.
One--meson--exchange models have been used to describe one-- and two--nucleon induced
decay processes, $\Lambda N\to nN$ and $\Lambda NN\to nNN$, in a 
%many--body scheme in nuclear matter
nuclear matter framework which has been 
adapted for finite nuclei predictions via the local density approximation.
Besides, an intranuclear cascade code based on Monte Carlo techniques 
has simulated the final state interactions of the outgoing nucleons with the
residual nucleus. 

Unlike previous papers, here we have adopted a microscopic approach for 
describing the two--nucleon induced decay, which included the channels
$\Lambda nn\to nnn$ and $\Lambda pp\to npp$ in addition to the mode
$\Lambda np\to nnp$ already considered in previous phenomenological
studies. The use of a different theoretical approach for calculation meets our
main purpose of making the extraction of the ratio $\Gamma_n/\Gamma_p$ from data
less model dependent.

The results have been compared with previous finite
nucleus analyses \cite{prl,prc}, with special consideration for the two--nucleon 
induced decay mechanism, and with a considerable amount of recent data by 
KEK \cite{Ki02,Ok04,bhang,OutaVa,Kang05,Kim06}.
Apart from some difference ascribable to the phase space dependence and to the
different weak decay models adopted for the two--nucleon induced decay channels,
the present predictions for the observable ratios $N_n/N_p$ and $N_{nn}/N_{np}$ confirm
the finite nucleus results.

The single--neutron spectra for $^{12}_\Lambda$C 
and $^{89}_\Lambda$Y measured by KEK--E369 and KEK--E508
have been reproduced with reasonable accuracy.
On the contrary, the KEK--E508 single--proton spectrum for $^{12}_\Lambda$C
is in strong disagreement with our prediction. The theoretical distribution would
reproduce these data only by enforcing artificially large values for $\Gamma_n/\Gamma_p$,
which, in turn, would lead to a serious overestimation of the mentioned single--neutron spectra.
%Since our calculation is able to reproduce previous KEK--E307 single--proton data, 
As a consequence of that, through
the weak interaction model independent analysis of $N_n/N_p$ data for 
$^{12}_\Lambda$C we determined $\Gamma_n/\Gamma_p=0.95 \pm 0.21$, a value that 
largely overestimates pure theoretical predictions ranging between 0.3 and 0.5. 

The contribution of the two--nucleon induced decay channels in these analyses
turn out to be of moderate size when high nucleon kinetic energy cuts
such as 60 MeV are imposed. On the contrary, FSI effects reveal to be
of great importance when determining $\Gamma_n/\Gamma_p$ from 
single--nucleon data. A demonstration of that is the inapplicability of the 
relation $\Gamma_n/\Gamma_p=(N_n/N_p-1)/2$, which neglects FSI and two--nucleon
stimulated decay effects, sometimes used in experimental analyses.
Even the na\"if model of FSI proposed by Kim et al. in Ref.~\cite{Ki02}
is far from being supported by a detailed calculation.

Concerning nucleon--nucleon correlation observables, 
while there is fair agreement among our predictions and data
in the case of $N_{nn}$, we overestimate the observations in the case of $N_{np}$.
The origin of such a difference
is likely the same as the one responsible for the mentioned disagreement on the
single--proton spectra. Another indication supporting this
possibility comes from the fact that we also overestimate the experimental 
distributions for $N_{pp}$. We have thus advanced the hypothesis that 
the experiment could have systematically underestimated the number of protons emitted in
$^{12}_\Lambda$C decay, thus leading
to the underestimation of $N_p$, $N_{np}$ and $N_{pp}$ spectra.
On the contrary, for the observables involving only neutron detection, $N_n$ and $N_{nn}$,
theory and experiment turn out to agree. The same conclusion
can be drawn in terms of the finite nucleus results.
A clarification of the mentioned eventuality is desirable for an accurate determination
of the $\Gamma_n/\Gamma_p$ ratio.

Indeed, in this paper as well as in previous ones \cite{prl,prc},
$\Gamma_n/\Gamma_p$ has been determined from data on $N_{nn}/N_{np}$.
Here, for $^{12}_\Lambda$C
we have obtained $\Gamma_n/\Gamma_p=0.43\pm 0.10$, whose central value 
could be lowered to about 0.3 if the measured neutron--proton distributions
resembled those of our evaluations. The conclusion remains the same
by using the finite nucleus analysis. The effect of the two--nucleon induced 
decay modes on the extracted $\Gamma_n/\Gamma_p$ turn out to be small due
to the restrictions imposed on angular and energy correlations.
In any case, one has to bear in mind that
ratios in the interval 0.3-0.5 would be compatible with most of 
today's pure theoretical evaluations, which are also affected by uncertainties,
especially due to the degree of arbitrariness in the experimentally unknown 
baryon couplings.
These models can also reproduce the observed total non--mesonic decay rates.

Before concluding, we want to make a brief comment on the two--nucleon 
induced mode. Because of the small or anyhow moderate dependence of the 
determined $\Gamma_n/\Gamma_p$ values on these decay channels, triple coincidence 
measurements reveal to be necessary for the purpose of disentangling one--
and two--nucleon stimulated decays in observed events.
%determining $\Gamma_2/\Gamma_1$.

To summarize, while we can safely assert that analyses of correlation measurements 
definitely solve the longstanding puzzle on the ratio $\Gamma_n/\Gamma_p$,
single--nucleon spectra studies still provide ratios which are
incompatible with what is obtained from pure theoretical models. We hope it will be possible
to clarify soon the reasons of such a discrepancy.
%It would be desirable to clarify as soon as possible the reasons of such 
%a discrepancy.

%%%%%%%%%%%%%%%%%%%%%%%%%%%%%
\section*{Acknowledgments}
%%%%%%%%%%%%%%%%%%%%%%%%%%%%%%%
This work is partly supported by EURIDICE HPRN-CT-2002-00311, 
contract FIS2005-03142 from MEC (Spain) and FEDER,
%contract MIUR 2001024324\_007, 
INFN--MEC collaboration agreement number 06--36,
and Generalitat de Catalunya contract 2005SGR-00343.
This research is part of the EU Integrated Infrastructure Initiative
Hadron Physics Project under contract number RII3-CT-2004-506078.

%%%%%%%%%%%%%%%%%%%%%%%%%%%%%%%


\begin{thebibliography}{100}
%%%%%%%%%%%%%%%%%%%%%%%%%%%%%%%

\bibitem{Al02}
W. M. Alberico and G. Garbarino, Phys. Rep. {\bf 369}, 1 (2002);
in {\em Hadron Physics}, IOS Press, Amsterdam, 2005, p.~125.
Edited by T. Bressani, A. Filippi and U. Wiedner.
Proceedings of the International School of Physics
``Enrico Fermi", Course CLVIII, Varenna (Italy), June 22 -- July 2, 2004.

\bibitem{Ra98}
E. Oset and A. Ramos, Prog. Part. Nucl. Phys. {\bf 41}, 191--253 (1998).

\bibitem{Ha01}
O. Hashimoto et al., Phys. Rev. Lett. {\bf 88}, 042503 (2002).

\bibitem{Sa05}
Y. Sato et al., Phys. Rev. {\bf C 71}, 025203 (2005).

\bibitem{Ki02}
J. H. Kim el al., Phys. Rev. {\bf C 68}, 065201 (2003).

\bibitem{Ok04}
S. Okada et al., Phys. Lett. {\bf B 597}, 249 (2004).

\bibitem{bhang}
H. Bhang, in {\em DAPHNE 2004: Physics at meson factories},
Frascati Phys. Ser. {\bf 36}, 243 (2005).
Edited by F. Anulli, M. Bertani, G. Capon, C. Curceanu--Petrascu,
F. L. Fabbri and S. Miscetti.
%(INFN, Laboratori Nazionali di Frascati, Frascati, Italy, 2005).

\bibitem{OutaVa}
H. Outa, in {\em Hadron Physics} (Ref.~\cite{Al02}) p.~219;
H. Outa et al., Nucl. Phys. {\bf A 754}, 157c (2005).
% S. Okada et al., Nucl. Phys. {\bf A 752}, 196c (2005).

\bibitem{Kang05}
B. H. Kang et al.,  Phys. Rev. Lett. {\bf 96}, 062301 (2006).
%nucl-ex/0509015.

\bibitem{Kim06}
M. J. Kim et al., nucl-ex/0601029.

\bibitem{fi}
M. Agnello et al., Nucl. Phys. {\bf A 754}, 399c (2005);
T. Bressani, E. Botta, A. Feliciello and V. Paticchio,
Nucl. Phys. {\bf A 754}, 410c (2005).

\bibitem{jparc}
T. Nagae, Nucl. Phys. {\bf A 754}, 443c (2005).

\bibitem{hyphi}
I. Tanihata et al., Hypernuclei with Stable Heavy Ion Beam and RI--beam Induced
Reactions at GSI (HypHI), Letter of Intent, February 2005.

\bibitem{prl}
G. Garbarino, A. Parre\~no and A. Ramos, Phys. Rev. Lett. {\bf 91}, 112501 (2003).

\bibitem{prc}
G. Garbarino, A. Parre\~no and A. Ramos, Phys. Rev. {\bf C 69}, 054603 (2004);
W. M. Alberico, G. Garbarino, A. Parre\~no and A. Ramos, nucl-th/0407046, in
{\em DAPHNE 2004: Physics at meson factories},
Frascati Phys. Ser. {\bf 36}, 249 (2005).
Edited by F. Anulli, M. Bertani, G. Capon, C. Curceanu--Petrascu,
F. L. Fabbri and S. Miscetti.

\bibitem{Os01}
D. Jido, E. Oset and J. E. Palomar, Nucl. Phys. {\bf A 694}, 525 (2001).

\bibitem{Pa02}
A. Parre\~{n}o and A. Ramos, Phys. Rev. {\bf C 65}, 015204 (2002);
A. Parre\~{n}o, A. Ramos and C. Bennhold, Phys. Rev. {\bf C 56}, 339 (1997).

\bibitem{It02} K. Itonaga, T. Ueda and T. Motoba,
Phys. Rev. {\bf C 65}, 034617 (2002).

\bibitem{Ba03}
E. Bauer and F. Krmpoti\'c, Nucl. Phys. {\bf A 717}, 217 (2003).

\bibitem{Ba04}
E. Bauer and F. Krmpoti\'c, Nucl. Phys. {\bf A 739}, 109 (2004).

\bibitem{Aj00}
S. Ajimura et al., Phys. Rev. Lett. {\bf 84}, 4052 (2000).

\bibitem{Ma04}
T. Maruta et al., nucl-ex/0509016;
Nucl. Phys. {\bf A 754}, 168c (2005).

\bibitem{asyth}
W.M. Alberico, G. Garbarino, A. Parre\~no and A. Ramos,
Phys. Rev. Lett. {\bf 94}, 082501 (2005).

\bibitem{barbero}
C. Barbero, A. P. Gale\~ao and F. Krmpoti\'c, Phys. Rev. {\bf C 72}, 035210
(2005); C. Barbero, C. De Conti, A. P. Gale\~ao and F. Krmpoti\'c, 
Nucl. Phys. {\bf A 726}, 267 (2003).

\bibitem{oka}
K. Sasaki, M. Izaki, and M. Oka, Phys. Rev. {\bf C 71}, 035502 (2005);
K. Sasaki, T. Inoue and M. Oka, Nucl. Phys. {\bf A 707}, 477 (2002).

\bibitem{assum}
A. Parre\~no, C. Bennhold and B.R. Holstein, Phys. Rev. {\bf C 70}, 
051601(R) (2004); Nucl. Phys. {\bf A 754}, 127c (2005).

\bibitem{Mo74}
A. Montwill et al., Nucl. Phys. {\bf A 234}, 413 (1974).

\bibitem{Sa91} A. Sakaguchi {\sl et al.}, Phys. Rev. {\bf C 43} (1991) 73.

\bibitem{Sz91}
J. J. Szymanski et al., Phys. Rev. {\bf C 43}, 849 (1991).

\bibitem{No95} H. Noumi et al., Phys. Rev. {\bf C 52}, 2936 (1995).

\bibitem{No95a}
H. Noumi et al., Proceedings of the {\em IV International
Symposium on Weak and Electromagnetic Interactions in Nuclei}, edited by
H. Ejiri, T. Kishimoto and T. Sato (World Scientific, Singapore, 1995), p.~550.

\bibitem{Ra97} A. Ramos, M. J. Vicente-Vacas and E. Oset,
Phys. Rev. {\bf C 55}, 735 (1997); {\bf 66}, 039903(E) (2002).

\bibitem{Ra94}
A. Ramos, E. Oset and L. L. Salcedo, Phys. Rev. {\bf C 50}, 2314 (1994).

\bibitem{Al00}
W. M. Alberico, A. De Pace, G. Garbarino and A. Ramos,
Phys. Rev. {\bf C 61}, 044314 (2000).

\bibitem{Os85}
E. Oset and L. L. Salcedo, Nucl. Phys. {\bf A 443}, 704 (1985).

\bibitem{Al00b}
 W. M. Alberico, A. De Pace, G. Garbarino, R. Cenni,
Nucl. Phys. {\bf A 668}, 113 (2000).

\bibitem{ba98}
E. Bauer, A. Polls and A. Ramos,
Phys. Rev. {\bf C 58}, 1052 (1998).

\bibitem{Al84}
W. M. Alberico, M. Ericson and A. Molinari, Ann. Phys. {\bf C 154}, 356 (1984).

\bibitem{Ma87}
R. Machleidt, K. Holinde and Ch. Elster;
Phys. Rep. {\bf 149}, 1 (1987).

\bibitem{Br96}
M. B. Barbaro, A. De Pace, T. W. Donnelly and A. Molinari,
Nucl. Phys. {\bf A 596}, 553 (1996).


%\bibitem{Xi98}
%C. Xiangzhou, F. Jung, S. Wenqing, M. Yugang, W. Jiansong and Y. Wei,
%Phys. Rev. {\bf C 58}, 572 (1998).

\bibitem{Cu03}
J. Cugnon and P. Henrotte, Eur. Phys. J. {\bf A 16}, 393 (2003).

\end{thebibliography}
\end{document}